\documentclass[a4paper, 11pt]{article}

\usepackage{amsmath}
\usepackage{graphicx} 
\usepackage[margin=1in]{geometry}
\usepackage{tikz}
\usetikzlibrary{calc,math}
\usepackage{pgfplots}
\usepgfplotslibrary{dateplot}
\usepackage{eurosym}
\usepackage{natbib}
\usepackage{url}
\usepackage{comment}
\usepackage{booktabs}
\usepackage{makecell}
\usepackage{csvsimple}
\usepackage{subcaption}
\usepackage[dvipsnames]{xcolor}
\colorlet{UrlBlue}{MidnightBlue!70!black}
\colorlet{darkgreen}{green!60!black}
\usepackage[
  unicode=true,
  colorlinks=true,
  linkcolor=MidnightBlue,
  citecolor=OliveGreen,
  urlcolor=UrlBlue,
  filecolor=Magenta,
  bookmarksopen=true,
  bookmarksnumbered=true
]{hyperref}
\usepackage[capitalise,noabbrev]{cleveref}

\definecolor{bluegreen}{HTML}{2C5A5D}
\definecolor{turkis}{HTML}{559e8f}
\definecolor{lichtblau}{HTML}{12a7e7}
\definecolor{wifored}{HTML}{C3423F}

\pgfplotsset{compat=1.18}
\newcommand{\cotwo}{$\mathrm{CO}_2$}
\newcommand{\widebar}[1]{\mkern0mu\overline{\mkern0.7mu#1\mkern0.7mu}\mkern0mu}

\newcommand\blfootnote[1]{
  \begingroup
  \renewcommand\thefootnote{}\footnote{#1}
  \addtocounter{footnote}{-1}
  \endgroup
}

\begin{document}

\title{A Market Design Proposal for Decoupling Carbon and Electricity Prices}

\author{Simon Finster\footnote{Department of Economics, Johannes Kepler University Linz; Austrian Institute of Economic Research (WIFO); Email address: \href{mailto:simon.finster@jku.at}{simon.finster@jku.at}}
\and
Bernhard Kasberger\footnote{Department of Economics, Johannes Kepler University Linz; Austrian Institute of Economic Research (WIFO); Email address: \href{mailto:bernhard.kasberger@jku.at}{bernhard.kasberger@jku.at}}
\and
Simon Rütten\footnote{Austrian Institute of Economic Research (WIFO); Email address: \href{mailto:simon.ruetten@wifo.ac.at}{simon.ruetten@wifo.ac.at}}}

\date{\vspace{10pt}\today}

\maketitle

\begin{abstract}
    In European day-ahead electricity markets, carbon allowance costs passed through by marginal fossil plants raise consumer expenditure and generate inframarginal rents for non-emitting generators. We propose a settlement modification: when the zonal day-ahead price exceeds a threshold, non-emitting generation is remunerated at the clearing price minus a fixed CO$_2$-proxy deduction, while all other units continue to receive the uniform price. The mechanism thus reallocates a part of the inframarginal rents to consumers. Using hourly data we estimate static average expenditure reductions of about 8.5\% in Austria and 4.7\% in Germany in 2025. We discuss bidding incentives around the threshold, interactions with Contracts for Difference, implementation in coupled bidding zones, and a gas-cost variant for the 2022 energy crisis.
\end{abstract}
\blfootnote{\emph{Acknowledgments}: We thank Gabriel Felbermayr for helpful comments.}

\section{Introduction}
\label{sec:intro}

This article proposes a modification to the wholesale electricity market design with the intention to attenuate the link between \cotwo\ prices and electricity expenditures. Empirical research indicates that emissions costs are substantially passed through to wholesale prices in auction-based electricity markets \citep{FabraReguant2014}. Therefore, we introduce the conditional deduction of a \emph{\cotwo\ price proxy} that is applied to the remuneration of inframarginal eligible non-fossil generators (wind, solar, hydro, nuclear, geothermal, and biomass) during high-price periods. Concretely, when the day-ahead price exceeds a pre-specified threshold (e.g., 100~\euro/MWh), a fixed proxy amount (e.g., 28~\euro/MWh) is deducted from the uniform market-clearing price paid to eligible non-fossil generators. The deduction is intended to approximate the embedded \cotwo\ cost component in the wholesale price and can be periodically updated to reflect prevailing allowance prices. To reduce undesirable bidding incentives around the activation point, the deduction can be phased in over a price interval rather than activated at a single cutoff, thereby eliminating the payoff discontinuity that would otherwise create incentives for price bunching near the threshold (see \cref{sec:incentives}). The aim is a transfer effect from inframarginal rents of eligible non-fossil technologies to consumers, while largely preserving marginal dispatch incentives and the allocative efficiency properties of a uniform-price auction.

Carbon prices exist to internalize the negative externality of burning fossil fuels, which generates \cotwo\ emissions and adversely affects the climate. A standard efficiency-oriented policy response is to cap total \cotwo\ emissions and allocate tradable emission certificates to firms \citep{Schmalensee-Stavins-Lessons-Learned}. Such a cap-and-trade system induces a market price for emission allowances. In theory, this price equals the marginal cost of abatement for the last unit required to meet the cap, and thereby provides decentralized incentives to undertake relatively low-cost emission reductions first.

\begin{figure}
    \centering
\begin{tikzpicture}
    \begin{axis}[
        date coordinates in=x,
        xmin=2019-01-01,
        xmax=2025-12-15,
        xlabel={},
        xtick={2019-01-01, 2021-01-01, 2023-01-01, 2025-01-01},
        xticklabel={1 Jan \year},
        ylabel={Carbon price (\euro/t)},
        grid=none,
        width=12cm,
        height=7cm,
        unbounded coords=discard,
        legend pos=north west,
        legend style={font=\small},
    ]
        \addplot[
            bluegreen,
            thick,
            no marks,
            solid
        ] table [
            x=ts_hour,
            y=carbon_price_eu,
            col sep=comma
        ] {figures/carbon_price_all.csv};
        \addlegendentry{EU}

        \addplot[
            lichtblau,
            thick,
            no marks,
            dotted
        ] table [
            x=ts_hour,
            y=carbon_price_cali,
            col sep=comma
        ] {figures/carbon_price_all.csv};
        \addlegendentry{California}

        \addplot[
            wifored,
            thick,
            no marks,
            dashed
        ] table [
            x=ts_hour,
            y=carbon_price_china,
            col sep=comma
        ] {figures/carbon_price_all.csv};
        \addlegendentry{China}

    \end{axis}
\end{tikzpicture}
    \caption{Carbon prices in major emissions trading systems}
    \label{fig:co2_price}
    \vspace{10pt}
    \begin{minipage}{0.9\textwidth}
        \footnotesize{\emph{Note: } Data represents daily allowance prices in the EU ETS, Chinese National ETS, and Californian Cap-and-Trade Program. Missing values (NAs) are linearly approximated. Data source: International Carbon Action Partnership.}
    \end{minipage}
\end{figure}
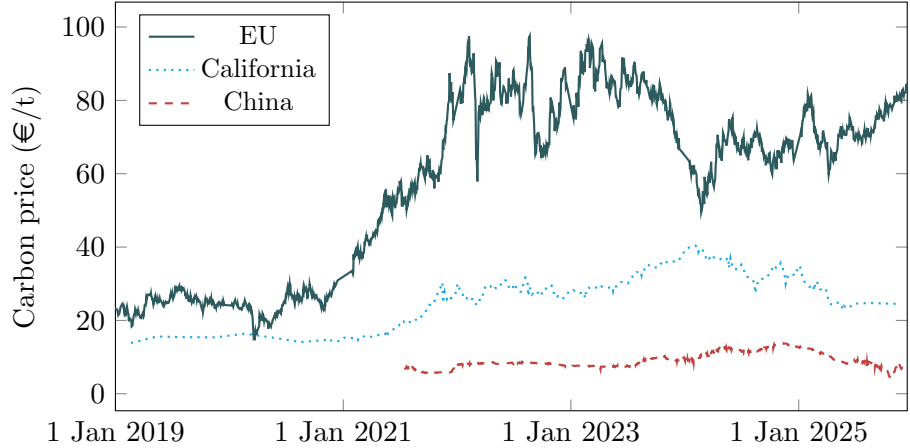

The European Union introduced the European Union Emissions Trading System (EU ETS) for carbon dioxide in 2005.\footnote{Other regions with cap-and-trade systems for \cotwo\ include California and Quebec, China, the United Kingdom, Switzerland, and South Korea.} Electricity and heat generators, as well as a broad set of energy-intensive industrial installations, are required to surrender \cotwo\ allowances for their emissions. \cref{fig:co2_price} shows the evolution of allowance prices across three major emission trading systems. In 2019, the EU ETS allowance price stood at approximately 20~\euro/t~\cotwo\ and has since risen to about 80~\euro/t by 2025, reflecting the tightening of allowance supply and growing policy ambition. California's Cap-and-Trade Program has seen a more moderate increase over the same period, with prices rising gradually from roughly 15~\euro/t, peaking in 2024 before retreating slightly. Despite trading at an even lower price level, remaining below 15~\euro/t throughout its existence, China's recent national ETS followed a similar trajectory, peaking in 2024, before decreasing slightly in 2025. The EU ETS has thus emerged as the most stringent and highest-priced carbon market among the three and \cotwo\ prices have become an important driver of production costs in the power sector and other covered industries.

Because gas- and coal-fired power plants must surrender allowances for each ton of \cotwo\ they emit, the \cotwo\ price enters their marginal production costs, whereas the marginal costs of renewable generators remain essentially unaffected. As a result, the allowance price tends to shift the merit order toward low- and zero-carbon technologies and to raise the wholesale electricity price in all hours in which fossil plants are marginal. Since gas-fired plants operate only if their expected revenues exceed fuel, carbon allowance, and operating costs, a higher \cotwo\ price increases the electricity price whenever gas (or coal) units set the clearing price.

On the demand side, the pass-through of carbon prices into electricity prices decreases incentives for substitution to (low-carbon) electric technologies outside the power sector, for example electric vehicles. If higher \cotwo\ prices substantially raise wholesale electricity prices, the relative price of (relatively clean) electricity compared to directly burned fossil fuels (gas, oil, coal) may change only modestly. In that case, carbon pricing strengthens incentives to switch from coal to gas or renewables \emph{within} the power sector, but may provide weak incentives for households and firms to substitute away from fossil-based end-use technologies. Put differently: the same policy that decreases the supply-side carbon intensity of electricity can---via wholesale-price pass-through---dampen demand-side electrification incentives because it raises, in conjunction with a uniform clearing price across technologies, inframarginal electricity remuneration.

\begin{figure}[htbp]
  \centering
  \includegraphics[scale=0.35]{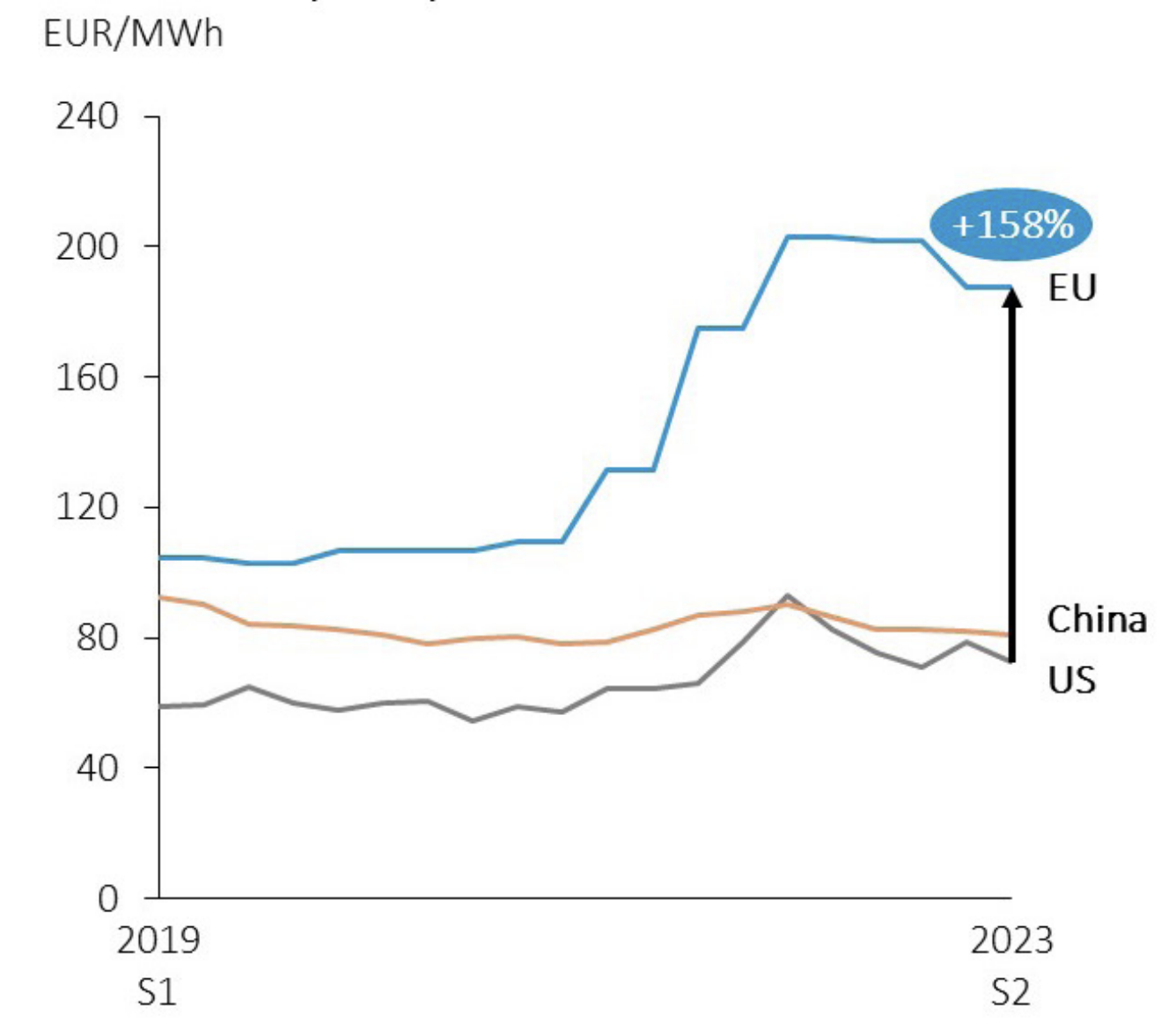}
  \caption{Industrial retail power prices}
  \label{fig:industrial_retail_power_prices}
  \vspace{10pt}
  \begin{minipage}{0.9\textwidth}
  \footnotesize
  \emph{Note:} Adapted from \citet{EC_Draghi_Report_Competitiveness}, The Draghi Report, Figure 1.
  \end{minipage}
\end{figure}

A second concern with carbon prices increasing electricity prices is competitiveness and carbon leakage. When electricity prices rise, energy-intensive and trade-exposed industries face stronger pressure to relocate production to jurisdictions with lower energy costs. Such relocation may leave global emissions largely unchanged---or even increase them if production shifts to less regulated regions outside the EU ETS---while imposing domestic costs in the form of lost output, employment, and industrial capability. From a climate perspective, shifting \cotwo-intensive production abroad can therefore represent displacement rather than decarbonization. These concerns are particularly salient in Europe, where energy prices have been persistently high in international comparison (see \cref{fig:industrial_retail_power_prices}). While multiple factors contribute to this gap---including fuel prices, network costs, and taxes---\cotwo\ pricing is a systematic component in hours when fossil generators set the marginal price.

Our market design proposal is motivated by this trade-off. The objective is to reduce the extent to which \cotwo\ costs translate into inframarginal rents for non-emitting electricity producers and into electricity expenditures for consumers during high-price periods. By attenuating the pass-through channel on the revenue side for inframarginal non-emitting technologies---while preserving the marginal cost signal for fossil plants---the \cotwo\ proxy deduction has two major advantages: (1) it strengthens electrification incentives, and (2) it alleviates pressure on industrial competitiveness and consumer budgets; and both are achieved without undermining the original goal of carbon pricing. 

There are academic and recent real-world discussions of the welfare effects of climate policies in the electricity sector. Academically, \citet{HirthUeckerdt2013} argue that carbon prices increase producer surplus and decrease consumer surplus (similar to our motivation) and show that subsidies to renewables can transfer welfare from producers to consumers (via the merit order effect \citep{Sensfuss-2008}). In the policy domain, the Italian government suggested in early 2026 that consumers pay the fossil power plants' \cotwo\ costs, which would reduce their marginal costs and lower electricity prices \citep{FT-Italy}. This would lower the non-emitting producers' profits and change the merit order in favor of emitting producers. In contrast, our proposal does not change the merit order in favor of emitting producers while also lowering consumers' electricity expenditures. 

\Cref{sec:impact-of-co2-on-wholesale-electricity-price} discusses the impact of \cotwo-prices on electricity prices. We present the proposed market design in \cref{sec:market-design-proposal} and quantify the impact on electricity expenditure. \Cref{sec:incentives} provides an analysis of how the settlement rule impacts the incentives in the day-ahead auction. We discuss implementation and institutional details in \cref{sec:discussion}. Finally, \cref{sec:application-to-gas} discusses how a related mechanism could be used for ``decoupling'' gas and electricity prices during extreme gas price hikes.

\section{The Impact of \cotwo-Prices on Electricity Prices}
\label{sec:impact-of-co2-on-wholesale-electricity-price}

In this section, we discuss how \cotwo\ prices affect wholesale electricity prices and the distribution of surplus between producers and consumers. To build intuition, we begin with a stylized single-period benchmark under perfect competition. We then discuss how intertemporal constraints (e.g., start-up costs, ramping, storage, block bids) and imperfect competition can alter price formation and pass-through in practice.

We consider a wholesale electricity market in which prices are determined by several uniform-price auctions, one for each time period. In each auction, suppliers submit offer schedules (bid supply curves) and wholesale buyers (retailers and large industrial consumers) submit demand schedules. The market-clearing price equates aggregate supply and demand in the respective period, and all accepted quantities are settled at this uniform price. Across periods, bids and operational decisions may be independent or linked through intertemporal constraints and bidding formats. For example, the European day-ahead market determines electricity prices for each 15-minute interval of the following day. The auction allows ``block bids'' that link intervals, introducing non-convexities that we ignore in the single-period benchmark.

A relevant concept is the \emph{pass-through} of \cotwo\ costs into wholesale electricity prices. Pass-through describes how much of a change in marginal emissions costs faced by fossil generators is reflected in the market-clearing electricity price. Formally, letting $p_t$ denote the wholesale price and letting $m_t$ denote the per-MWh emissions-cost component of the marginal technology in period $t$, the pass-through rate can be defined as $\partial p_t/\partial m_t$. A pass-through of one means that a one-euro increase in marginal emissions cost raises the wholesale price by one euro whenever a fossil unit is marginal. A pass-through below one means that prices rise by less than the cost increase. A pass-through above one means that prices rise by more than the increase in marginal emissions cost, so that the price response amplifies the underlying cost shock. Since marginal pricing applies the market-clearing price to all accepted output, the degree of pass-through determines both the expenditure impact on consumers and the extent to which higher \cotwo\ prices translate into inframarginal rents for non-emitting generators.

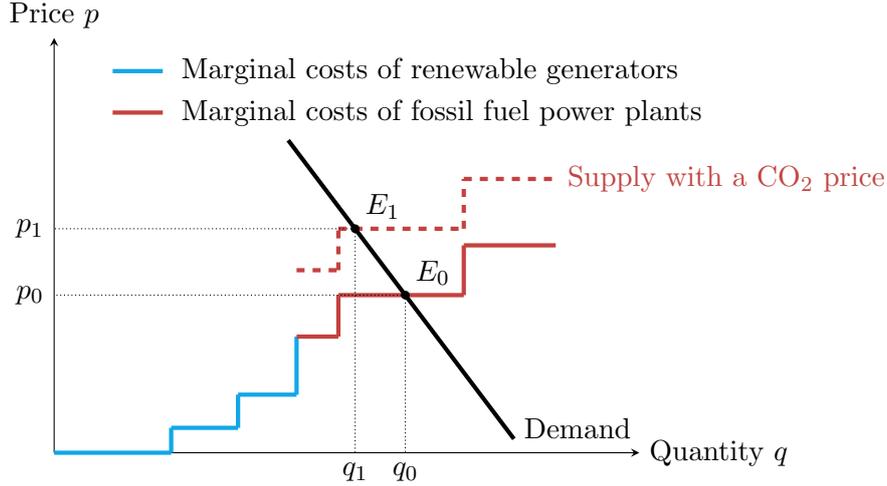
\begin{figure}
    \centering
    \begin{tikzpicture}[scale=1.1, >=stealth]

  \draw[->] (0,0) -- (7,0) node[right] {Quantity $q$};
  \draw[->] (0,0) -- (0,5) node[above] {Price $p$};

  \draw[ultra thick] (2.8,3.76667) -- (5.5,0.166667);
  \node [right]at (5.5,0.3){Demand};


  \draw[ultra thick, lichtblau] (0,0) -- (1.4,0);
  \draw[ultra thick, lichtblau] (1.4,0) -- (1.4,0.3);
  \draw[ultra thick, lichtblau] (1.4,0.3) -- (2.2,0.3);
  \draw[ultra thick, lichtblau] (2.2,0.3) -- (2.2,0.7);
  \draw[ultra thick, lichtblau] (2.2,0.7) -- (2.9,0.7);
  \draw[ultra thick, lichtblau] (2.9,0.7) -- (2.9,1.4);

  \draw[ultra thick,wifored] (2.9,1.4) -- (3.4,1.4);
  \draw[ultra thick,wifored] (3.4,1.4) -- (3.4,1.9);
  \draw[ultra thick,wifored] (3.4,1.9) -- (4.9,1.9);
  \draw[ultra thick,wifored] (4.9,1.9) -- (4.9,2.5);
  \draw[ultra thick,wifored] (4.9,2.5) -- (6.0,2.5);


  \draw[ultra thick,wifored,dashed] (2.9,2.2) -- (3.4,2.2);
  \draw[ultra thick,wifored,dashed] (3.4,2.2) -- (3.4,2.7);
  \draw[ultra thick,wifored,dashed] (3.4,2.7) -- (4.9,2.7);
  \draw[ultra thick,wifored,dashed] (4.9,2.7) -- (4.9,3.3);
  \draw[ultra thick,wifored,dashed] (4.9,3.3) -- (6.0,3.3)
    node[right] {Supply with a \cotwo\ price};



  \coordinate (Q0) at (4.2,1.9);
  \fill (Q0) circle (1.5pt) node[above right] {$E_0$};
  \draw[densely dotted] (Q0) -- (4.2,0) node[below] {$q_0$};
  \draw[densely dotted] (Q0) -- (0,1.9) node[left] {$p_0$};

  \coordinate (Q1) at (3.6,2.7);
  \fill (Q1) circle (1.5pt) node[above right] {$E_1$};
  \draw[densely dotted] (Q1) -- (3.6,0) node[below] {$q_1$};
  \draw[densely dotted] (Q1) -- (0,2.7) node[left] {$p_1$};

  \draw[lichtblau,ultra thick] (0.7,4.6) -- (1.3,4.6);
  \node[anchor=west] at (1.4,4.6) {Marginal costs of renewable generators};

  \draw[wifored,ultra thick] (0.7,4.1) -- (1.3,4.1);
  \node[anchor=west] at (1.4,4.1) {Marginal costs of fossil fuel power plants};
\end{tikzpicture}
    \caption{Equilibrium effect with \cotwo-prices}
    \label{fig:merit_order}
    \vspace{10pt}
    \begin{minipage}{0.9\textwidth}
        \footnotesize\emph{Note:} The \cotwo-price increases the marginal costs of fossil fuel power plants. Supply curves respond to an increase in marginal costs. In the new equilibrium, the price is higher and demand lower when fossil fuel power plants are active.
    \end{minipage}
\end{figure}

To illustrate the price impact of \cotwo\ prices in a simple setting, we assume perfect competition and a single-period market (no intertemporal linkages in bidding), so that bids equal marginal costs and the supply curve coincides with the merit order. \Cref{fig:merit_order} then illustrates supply and demand in the electricity market. Without a \cotwo\ price, the market clears at price~$p_0$ and quantity~$q_0$ (point~$E_0$). Introducing a \cotwo\ price raises the marginal costs of fossil-fired plants but leaves those of renewable generators unchanged, shifting the fossil supply curve upward. In this benchmark, the pass-through of marginal emissions costs is one whenever a fossil unit is price-setting. Whenever a fossil-fired plant is marginal (as depicted), the new equilibrium features a higher price~$p_1 > p_0$ and lower quantity~$q_1 < q_0$, reflecting both the cost increase for carbon-intensive technologies and the demand response to higher prices. Because all inframarginal generators receive the higher clearing price, the producer surplus of non-emitting technologies rises, redistributing welfare from consumers to producers in high-price periods (see also, e.g.,~\citet{Sijm01012006} and \citet{HirthUeckerdt2013}).

Intertemporal linkages can weaken the one-period merit-order logic. In hours in which no emitting unit is marginal, an increase in the \cotwo\ price does not mechanically raise the marginal cost of the price-setting technology, and the direct pass-through channel is therefore absent. However, electricity supply is not fully separable across hours. Many thermal plants face non-convex operating constraints---such as start-up costs, minimum stable generation levels, and ramping limits---and market designs allow intertemporal bids (e.g., block bids) that link acceptance decisions across periods. In such settings, a plant may be committed in low-price hours as part of an intertemporal schedule even when the day-ahead price in those hours is below its short-run marginal cost. Losses incurred in these hours must then be recovered in other hours of operation in order for the plant (or the linked bid) to break even. An increase in the \cotwo\ price raises the variable cost of emitting units in all hours in which they operate and can therefore increase the required scarcity rents in the hours that finance these commitments. As a result, \cotwo\ price changes can affect electricity prices even in hours in which emitting units are not marginal, and price responses can be amplified in peak hours if these hours are the margin on which intertemporal recovery takes place. These intertemporal linkages imply that the response of peak-hour prices to changes in \cotwo\ costs need not be one-for-one and can, in principle, exceed the contemporaneous increase in marginal emissions costs (pass-through above one).

Intertemporal effects can also operate across bidding zones through market coupling. For example, even if no domestic coal plant is running, interconnector flows may transmit scarcity conditions and cost shocks from neighboring zones; when imports are marginal, foreign units subject to higher \cotwo\ costs can influence domestic prices.

Imperfect competition can also modify cost pass-through because bids need not equal marginal costs. In a day-ahead auction, the market-clearing price can be decomposed as the marginal cost of the price-setting unit plus a markup that reflects the competitiveness of residual demand. When marginal costs increase (e.g., through higher \cotwo\ prices), the equilibrium markup may adjust: markups can fall if competition intensifies at higher cost levels or if demand becomes more elastic, implying pass-through below one. In that case, part of the cost increase is absorbed by producers through lower margins rather than being fully transmitted to consumers. More generally, the pass-through rate depends on market structure, demand elasticity, and the curvature of supply and residual demand \citep{Sijm-market-structure}. 

The extent to which carbon allowance costs are passed through to wholesale prices is therefore central for consumer expenditure and the distribution of surplus in electricity markets. Empirical evidence indicates that the pass-through of \cotwo\ costs into wholesale electricity prices in European markets is high. \citet{FabraReguant2014} study the Spanish day-ahead market during the first phase of the EU ETS using plant-level bid data and an instrumental-variables strategy.\footnote{\citet{FabraReguant2014} use detailed plant-level bid data for the Spanish wholesale market from 2004--2006, instrumenting marginal emissions costs with the allowance price. They estimate both reduced-form and structural models of bidding behavior.} They find that a one euro increase in emissions costs raises wholesale prices by more than eighty cents on average, with pass-through close to one during peak hours when dynamic constraints are less binding. Their structural analysis shows that generators fully internalize the opportunity cost of permits, and that the high pass-through reflects the combination of highly inelastic short-run demand, correlated cost shocks across firms, and limited price rigidities in the auction format. Changes in the \cotwo\ price therefore feed almost one-for-one into wholesale electricity prices, with direct consequences for the distribution of rents between carbon-intensive and low-carbon technologies. \citet{Hintermann2016} documents similarly high pass-through rates for Germany.

Finally, a back-of-the-envelope calculation illustrates the magnitude of the cost increases. Let $\eta$ denote the thermal efficiency of a fossil plant, $p_{\text{fuel}}$ the fuel price per MWh$_{\text{th}}$, and $e_{\text{fuel}}$ the emission intensity per MWh$_{\text{th}}$ of fuel burned. The marginal cost of electricity production is then approximately
\begin{equation}\label{eq:mc-fossil}
  \text{MC} \;\approx\; \frac{1}{\eta}\bigl(p_{\text{fuel}} + e_{\text{fuel}}\, p_{\text{CO}_2}\bigr),
\end{equation}
where $p_{\text{CO}_2}$ is the allowance price per ton. The term $e_{\text{fuel}}\,p_{\text{CO}_2}/\eta$ captures the \cotwo\ cost per MWh of electricity; its share in total marginal cost rises with both the emission intensity and the allowance price.

For a combined-cycle gas plant with $\eta = 0.55$ and $e_{\text{fuel}} = 0.2$~tCO$_2$/MWh$_{\text{th}}$, the implied emission intensity is approximately $0.36$~tCO$_2$/MWh$_{\text{el}}$. At a gas price of \euro\,30/MWh and a \cotwo\ price of \euro\,80/t, Equation~\eqref{eq:mc-fossil} yields a marginal cost of roughly \euro\,84/MWh, of which about one third is attributable to \cotwo\ costs. Doubling the allowance price raises the marginal cost to approximately \euro\,113/MWh, with \cotwo\ costs accounting for more than half.

For a lignite plant with $\eta = 0.40$ and $e_{\text{fuel}} = 0.4$~tCO$_2$/MWh$_{\text{th}}$, the emission intensity is $1.0$~tCO$_2$/MWh$_{\text{el}}$---nearly three times that of gas. At a lignite price of \euro\,10/MWh$_{\text{th}}$ and a \cotwo\ price of \euro\,80/t, the implied marginal cost is about \euro\,105/MWh, of which roughly three quarters are \cotwo\ costs. Doubling the allowance price would raise marginal costs to approximately \euro\,185/MWh, making \cotwo\ the dominant cost component.

\section{Market Design Proposal}
\label{sec:market-design-proposal}

This section proposes a mechanism that reduces the extent to which carbon costs translate into electricity expenditures during high-price periods.

\subsection{The Mechanism}

The key idea is a simple settlement rule in the electricity wholesale market: a conditional \emph{\cotwo\ proxy deduction} is applied only to the remuneration of eligible non-fossil generators (wind, solar, hydro, nuclear, geothermal, and biomass) in hours in which the uniform day-ahead price is high.

Let $p_t$ denote the uniform day-ahead market-clearing price in period $t$. Fix a price threshold $\widebar p$ (e.g., $100$~\euro/\text{MWh}) and a proxy amount $\delta_t \ge 0$ (e.g., $28$~\euro/\text{MWh}). Eligible non-fossil generators (wind, solar, hydro, nuclear, geothermal, and biomass) receive an adjusted remuneration
\[
\widetilde p_t = p_t - \mathbf{1}\{p_t \ge \widebar p\}\,\delta_t,
\]
while all other generators continue to receive the uniform price $p_t$. For example, if $p_t = 158$~\euro/\text{MWh}, $\widebar p = 100$~\euro/\text{MWh}, and $\delta_t = 28$~\euro/\text{MWh}, eligible generators receive $\widetilde p_t = 130$~\euro/\text{MWh}. Let $q_t$ denote total demand and $q_t^r$ the production of eligible generators. Total electricity expenditure in period $t$ is then
\begin{align}\label{equ:total-expenditure}
\text{Exp}_t = q_t^r \widetilde p_t + (q_t - q_t^r)p_t.
\end{align}
A wholesale consumer with demand $D_t$ incurs the payment $D_t\cdot\text{Exp}_t / q_t  $.

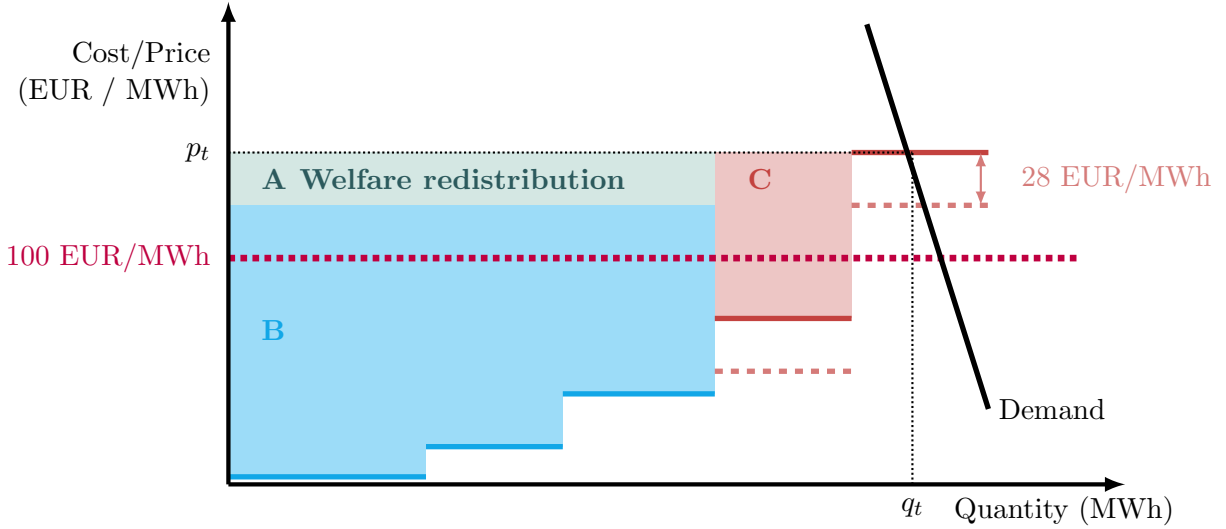
\begin{figure}[t]
\centering
    \begin{tikzpicture}[x=1cm,y=1cm,>=latex]
  \def\xmax{12.8}
  \def\ymax{7.2}

  \def\xR{7.4}    
  \def\xQ{10.0}   
  \def\y0{0.8}    
  \def\yP{5.2}    
  \def\yT{3.8}    
  \def\dCOtwo{0.7}

  \node[anchor=east] at (0.9,6.5) {Cost/Price};
  \node[anchor=east] at (0.9,6.0) {(EUR / MWh)};
  \node at (12.0,0.45) {Quantity (MWh)};

  \fill[cyan!35] (1,\y0+.1) rectangle (3.6,\yP-\dCOtwo);
  \fill[cyan!35] (3.6,1.3) rectangle (5.4,\yP-\dCOtwo);
  \fill[cyan!35] (5.4,2.0) rectangle (\xR,\yP-\dCOtwo);
  \fill[turkis!25] (1,\yP-\dCOtwo) rectangle (\xR,\yP);
  \fill[wifored!30] (\xR,3) rectangle (\xR+1.8,\yP);

  \draw[line width=2pt,lichtblau] (1,\y0+.1) -- (3.6,\y0+.1);
  \draw[line width=2pt,lichtblau] (3.6,1.3) -- (5.4,1.3);
  \draw[line width=2pt,lichtblau] (5.4,2.0) -- (\xR,2.0);

  \draw[line width=2pt,wifored] (\xR,3) -- (\xR+1.8,3);
  \draw[line width=2pt,wifored] (\xR+1.8,\yP) -- (\xQ+1,\yP);

    \draw[line width=2pt,wifored!70,dashed] (\xR,3-\dCOtwo) -- (\xR+1.8,3-\dCOtwo);
  \draw[line width=2pt,wifored!70,dashed] (\xR+1.8,\yP-\dCOtwo) -- (\xQ+1,\yP-\dCOtwo);

  
  \draw[<->,thick,wifored!70] (\xQ+0.9,\yP-\dCOtwo) -- (\xQ+0.9,\yP);
  \node[anchor=west,text=wifored!70] at (\xQ+1.3,\yP-0.35) {28 EUR/MWh};

  \draw[densely dotted,thick] (1,\yP) -- (\xQ,\yP);
  \draw[densely dotted,thick] (\xQ,\y0) -- (\xQ,\yP);

  \draw[line width=2.5pt,dotted, purple] (1,\yT) -- (\xmax-0.6,\yT);
  \node[anchor=east,text=purple] at (0.9,\yT) {100 EUR/MWh};

  \node[anchor=east] at (0.9,\yP) {$p_t$};
  \node[anchor=north] at (\xQ,\y0-0.05) {$q_t$};

  \node[anchor=west, text=bluegreen, font=\bfseries] at (1.8,4.85) {Welfare redistribution};
  \node[anchor=west, text=bluegreen, font=\bfseries] at (1.3,4.85) {A};

    \node[anchor=west, text=lichtblau, font=\bfseries] at (1.3,2.85) {B}; 
    \node[anchor=west, text=wifored, font=\bfseries] at (\xR+.3,4.85) {C}; 

  \draw[ultra thick,->] (1,\y0) -- (1,\ymax);
  \draw[ultra thick,->] (1,\y0) -- (\xmax,\y0);

  \draw[line width=2pt,black]
    (\xQ-0.6,\ymax-0.3) -- (\xQ+1.0,\y0+1.0);
    \node [right] at (\xQ+1.0,\y0+1.0) {Demand};

\end{tikzpicture}
\caption{Conditional \cotwo\ proxy deduction in high-price hours}
\label{fig:proxy_deduction}
\vspace{10pt}
\begin{minipage}{0.9\textwidth}
\footnotesize\emph{Note:} The figure illustrates a settlement rule under which eligible non-fossil generators receive $\widetilde p = p - \delta$ whenever the uniform day-ahead price $p$ exceeds a threshold (e.g., 100~\euro/MWh). The deducted amount $\delta$ proxies embedded \cotwo\ costs and is rebated to consumers. Under uniform marginal pricing (without the deduction), all inframarginal generators receive $p_t$; eligible producer surplus would correspond to areas $A+B$, and fossil producer surplus to area $C$. Under the proposed settlement rule, area $A$ is transferred from eligible producer surplus to consumer surplus. The blue area (B) represents the remaining producer surplus of eligible generators, the red area (C) is the producer surplus of fossil-based plants, and the green area (A) is the welfare gain of consumers and welfare loss of eligible generators.
\end{minipage}
\end{figure}

\cref{fig:proxy_deduction} illustrates the mechanism in a period where $p_t \ge \widebar p$ and a fossil unit is marginal. The mechanism implements a conditional redistribution of inframarginal rents from eligible non-fossil generators to consumers without altering the market-clearing price or dispatch outcome, while leaving fossil producer surplus unchanged (see figure for details).

The mechanism is not a revenue cap. Unlike a ceiling that limits total remuneration, the proxy deduction leaves eligible generators fully exposed to marginal price movements. Formally, $\partial \widetilde p_t / \partial p_t = 1$ for $p_t \ge \widebar p$: revenue increases one-for-one with the wholesale price above the threshold. The mechanism therefore preserves scarcity signals and high-price exposure while redistributing a fixed component of inframarginal rents.

\subsection{Parameter Design}
\label{sec:parameter-design}

The threshold $\widebar p$ should be set above the marginal cost of fossil generation excluding carbon costs (e.g., fuel and operating costs of gas plants). This ensures that the deduction activates primarily in hours when fossil plants are marginal and carbon allowance costs drive the high wholesale price. If the threshold were set too low, the mechanism would redistribute rents even in hours when \cotwo\ costs do not materially affect prices, weakening its economic rationale.

The proxy amount $\delta_t$ should reflect the carbon cost component embedded in the marginal fossil technology. A transparent specification ties the proxy directly to the \cotwo\ cost of electricity from an efficient gas (or coal) plant:
\begin{equation}\label{eq:delta}
    \delta_t = \alpha \cdot e_{\text{gas,el}} \cdot p_{\mathrm{CO}_2,t},
\end{equation}
where $e_{\text{gas,el}}$ denotes the reference emissions intensity (t\cotwo/MWh$_\text{el}$) and $\alpha \in [0,1]$ governs the degree of redistribution. Setting $\alpha=1$ implements full deduction of the estimated carbon cost component, while $\alpha < 1$ (e.g., $\alpha=0.8$) leaves a fraction of the carbon rent with eligible generators as an implicit subsidy. The specification automatically adjusts the deduction from eligible units to the up-to-date carbon allowance prices. The proxy amount $\delta_t$ might or might not be dependent on time $t$.

To avoid unintended distortions, parameters should satisfy
\begin{equation}\label{eq:delta-not-too-high}
\widebar p - \delta_t > \max_i c_i^{\text{eligible}},
\end{equation}
where $c_i^{\text{eligible}}$ denotes the marginal cost of eligible technologies. Otherwise, some eligible technologies could become temporarily unprofitable despite being cost-efficient, which would create allocative inefficiencies in the merit order. Storage is a special case: its opportunity cost reflects past charging prices, so the deduction can mechanically compress storage margins in high-price hours. Policymakers should therefore decide explicitly whether storage should be eligible and, if so, calibrate $\widebar p$ and $\delta_t$ accordingly.

To eliminate the payoff discontinuity at $\widebar p$---which, as we show in \cref{sec:incentives}, creates incentives for price bunching---the deduction schedule can be smoothed. Let $\underline p < \widebar p$ and let the deduction schedule $d(p_t)$ increase linearly between these two prices. The remuneration to eligible generators is then
\[
\widetilde p_t = p_t - d(p_t),
\]
where
\begin{equation}\label{eq:linear_ramp}
d(p_t)=
\begin{cases}
0, & p_t \le \underline p,\\
\delta_t \dfrac{p_t-\underline p}{\widebar p - \underline p}, & \underline p < p_t < \widebar p,\\
\delta_t, & p_t \ge \widebar p
\end{cases}
\end{equation}
where $\delta_t$ is given by \cref{eq:delta}. Under the linear ramp, the deduction increases continuously from zero to $\delta$. As a result, a marginal reduction in the clearing price no longer produces a discrete jump in eligible-generator profit; instead, the lower deduction and the lower price trade off smoothly, eliminating the region in which small supply shifts generate discontinuously large profit gains. We argue in \cref{sec:incentives-ramp} that this removes the threshold-bunching incentive. Other functional forms (e.g., concave phase-ins) can be used in place of a linear ramp.

\subsection{Quantification}
\label{sec:quant}

To assess the magnitude of the proposed \cotwo\ proxy deduction, we conduct a static accounting exercise using hourly data on day-ahead prices, load, and generation by technology. The objective is to quantify the mechanical redistribution implied by our settlement rule, abstracting from behavioral responses in demand, bidding, dispatch, or investment.

We construct an hourly panel each for the German and the Austrian price zone, respectively, by merging day-ahead prices with realized load and generation data; the data source is ENTSO-E. Eligible non-fossil generation is defined as wind (onshore and offshore), solar, hydro (run-of-river and reservoir), nuclear, geothermal, and biomass; pumped storage is excluded in the baseline quantification to guarantee that the inequality in~\eqref{eq:delta-not-too-high} holds for all eligible technologies. Let $p_t$ denote the observed day-ahead price in hour $t$, $L_t$ the realized load, and $R_t$ the eligible non-fossil generation. Status quo electricity expenditure in hour $t$ is given by
\[
\text{Exp}^{\text{base}}_t = p_t \cdot L_t.
\]
Under the proposed mechanism, a transfer is triggered whenever the price exceeds the threshold $\widebar p=100$ \euro/MWh. For each such hour, the redistributed amount is
\begin{equation}\label{eq:transfer}
T_t = \mathbf{1}\{p_t \ge \widebar p\} \cdot \delta \cdot R_t,
\end{equation}
where $\delta$ is the proxy deduction and equal to 28~\euro/MWh.\footnote{\Cref{tab:sensitivity} reports outcomes for alternative parameter values, and \cref{tab:linear_ramp_results} for a specification with a linear ramp.} The counterfactual electricity expenditure faced by consumers is then
\[
\text{Exp}^{\text{new}}_t = \text{Exp}^{\text{base}}_t - T_t.
\]
Aggregating across hours yields annual baseline expenditure and the counterfactual expenditure together with total redistributed inframarginal rents. Average electricity expenditure per MWh before and after the reform is computed by dividing total expenditure by total load. Importantly, quantities ($L_t$ and $R_t$) and prices ($p_t$) are held fixed throughout this exercise. The calculation therefore captures only the direct redistribution effect of our settlement rule, not any induced changes in dispatch, bidding strategies, demand, or investment behavior.

This static approach provides a transparent benchmark for the immediate mechanical redistribution under fixed market outcomes. In reality, lower effective consumer prices in high-price hours may stimulate demand, while reduced inframarginal rents could affect long-run investment incentives. These dynamic responses are intentionally left out in order to isolate the immediate expenditure channel targeted by the proposal. Our analysis in \cref{sec:incentives} suggests that bidders may have an incentive to alter their bidding functions in some time periods.

Applying the settlement rule to hourly data for 2025 yields economically meaningful redistribution magnitudes. In Austria, the total annual load amounts to 59.2~TWh and baseline electricity expenditure sums to approximately 6.18~billion~\euro. The proposed proxy deduction would redistribute about 528~million~\euro\ from inframarginal eligible non-fossil generation to consumers. Under the maintained assumption of fixed quantities and prices, average electricity expenditure falls from 104.4~\euro/MWh to 95.5~\euro/MWh, corresponding to a reduction of roughly 8.9~\euro/MWh, or about 8.5\%.

For Germany, with an annual load of 466~TWh and baseline expenditure of about 43.2~billion~\euro, the implied redistribution amounts to approximately 2.04~billion~\euro. The average expenditure decreases from 92.8~\euro/MWh to 88.4~\euro/MWh, a reduction of around 4.4~\euro/MWh, or approximately 4.7\%. 

\begin{figure}[htbp]
  \centering
  \includegraphics[width=0.9\textwidth]{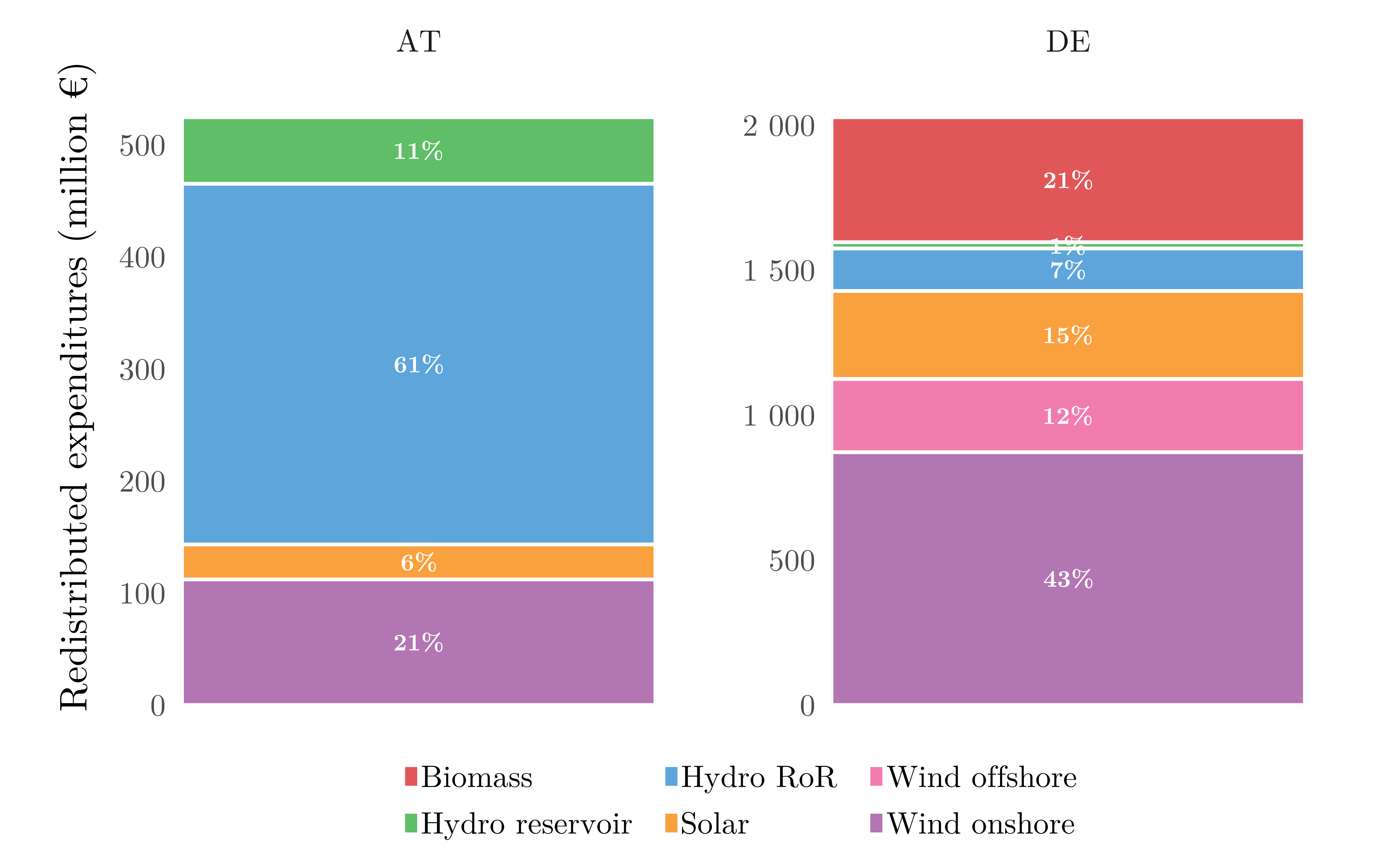}
  \caption{Redistributed expenditures by technology, Austria and Germany (2025).}
  \label{fig:p_tech_stack_no_ps}
  \vspace{10pt}
  \begin{minipage}{0.9\textwidth}
    \footnotesize\emph{Note: } The bars decompose the total redistribution implied by the proxy-deduction mechanism across eligible non-fossil technologies in Austria (AT) and Germany (DE). Heights are in million euros; labels within segments report each technology's share of the total redistribution in the respective country. Results are aggregated over all hours in 2025 using day-ahead prices and realized generation by technology. Note the different y-axis scales across panels. Technologies with less than 1\% contribution are left out for readability.
  \end{minipage}
\end{figure}

\Cref{fig:p_tech_stack_no_ps} decomposes the total redistribution implied by the proxy deduction across eligible non-fossil technologies for Austria and Germany. In Austria, the transfer is dominated by hydropower, reflecting the large share of run-of-river generation in high-price hours: run-of-river hydro accounts for about 61\% of the redistributed amount, followed by onshore wind (21\%) and reservoir hydro (11\%) while solar contributes around 6\%. In Germany, the redistribution is more diversified and driven primarily by wind: onshore wind represents the largest share (43\%), followed by biomass (21\%), solar (15\%), offshore wind (12\%), run-of-river hydro (7\%) and reservoir hydro (1\%). These patterns mirror the technology mix available in high-price hours and illustrate that the incidence of the mechanism across producers is system-specific. \Cref{tab:revenue_loss} shows the implied revenue losses by technology and zone resulting from the policy-induced redistribution in expenditure.

\begin{figure}[htbp]
  \centering
  \includegraphics[width=0.9\textwidth, trim=0 0 0 95, clip]{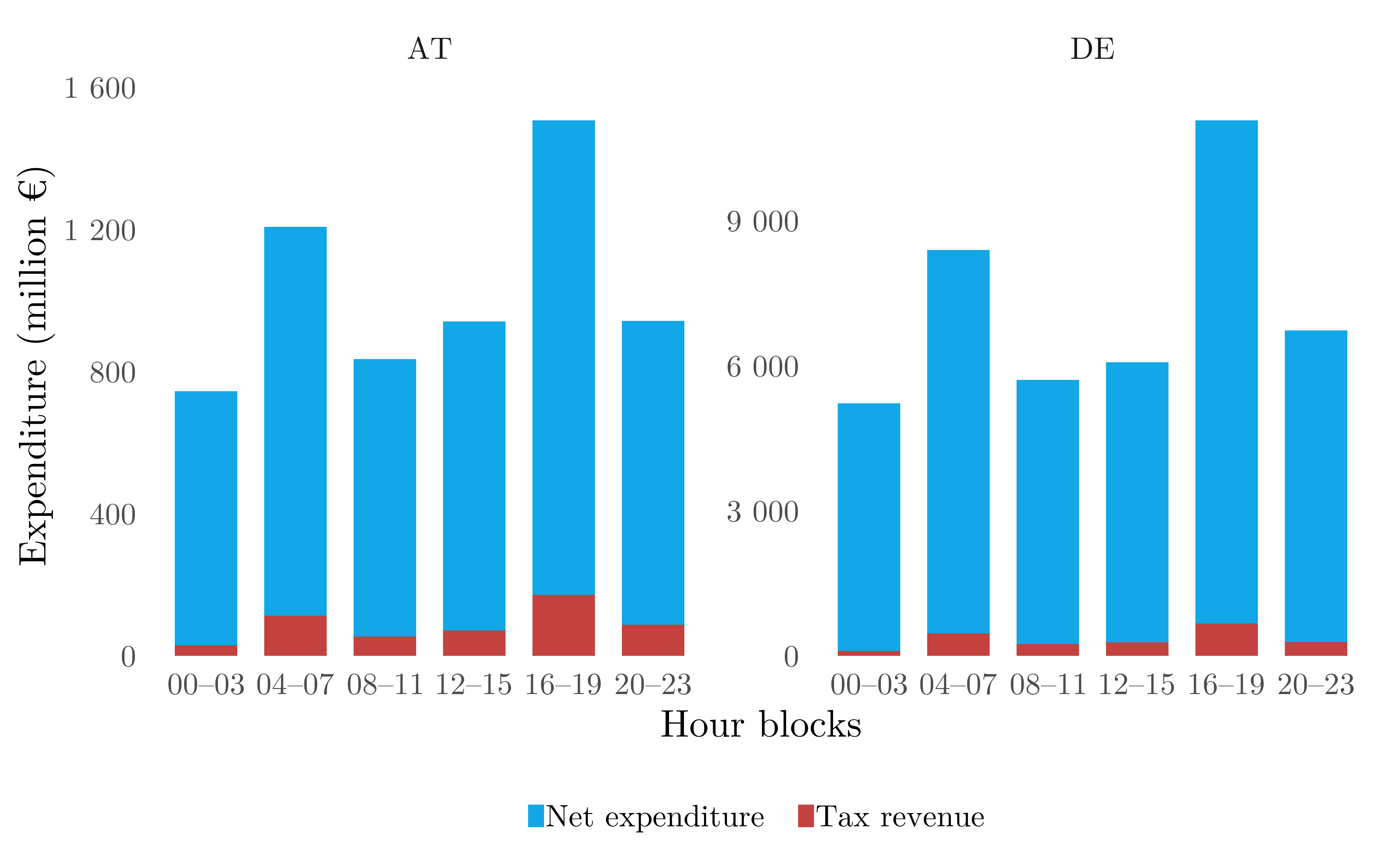}
  \caption{Electricity expenditure and redistribution by four-hour blocks for Austria (left) and Germany (right), 2025.}
  \label{fig:p_stack}
  \vspace{10pt}
  \begin{minipage}{0.9\textwidth}
    \footnotesize\emph{Note: } Blue bars show net consumer expenditure; red bars show the redistributed amount under the proposed proxy deduction. Note the different $y$-axis scales for readability.
  \end{minipage}
\end{figure}

Aggregating transfers by four-hour blocks reveals that the redistribution is concentrated in peak hours. In both Austria and Germany, the largest reductions in average electricity expenditure occur during late afternoon and early evening (4--7pm), when fossil generation is frequently marginal and wholesale prices highest (\cref{fig:p_stack}). In Austria, the average expenditure reduction in this block reaches 11.4\%, compared to around 6\% in Germany. Morning peak hours (4--7am) also show above-average reductions, while nighttime hours are largely unaffected.

This pattern is by design: the deduction activates when fossil units set the marginal price, concentrating the redistribution in peak-load periods rather than compressing the entire wholesale price distribution.

The average reduction in consumer expenditure per MWh implied by \cref{eq:transfer} is
\[ \,\delta \cdot \frac{\sum_{t:\,p_t \ge \widebar p} R_t}{\sum_t L_t},
\]
i.e., the proxy amount scaled by the share of eligible non-fossil generation in high-price hours relative to total load. This identity explains the cross-country differences: Austria exhibits a larger proportional reduction because a comparatively large volume of hydro generation is available in high-price hours---its renewable share exceeds Germany's by 18.3 percentage points in the 4--7pm block and by 16.1 percentage points in the 4--7am hours. In Germany, high-price periods coincide with lower renewable availability, and the larger load base further dilutes the per-MWh transfer.

\section{Incentives}
\label{sec:incentives}

\begin{figure}[htbp]
    \centering
    \includegraphics[width=0.85\textwidth]{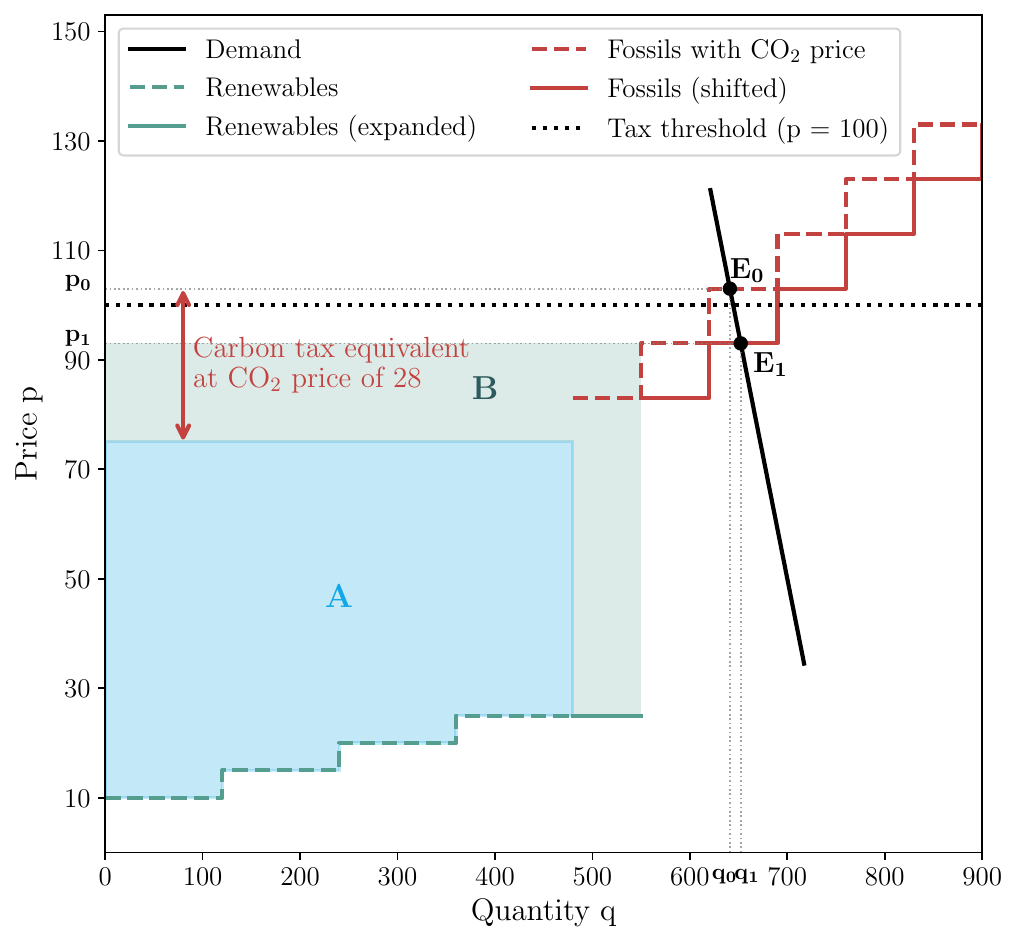}
    \caption{Incentives of renewables to expand demand around the threshold price.}
    \label{fig:threshold-incentives}
\end{figure}

The proposed mechanism creates a payoff discontinuity at the price threshold~$\widebar{p}$: when the market-clearing price lies above the threshold, non-emitting generators receive a reduced remuneration, but when it falls below, they retain the full market price. This discontinuity generates distinctive incentives for bidding in the day-ahead auction. We begin with the economic intuition illustrated by \cref{fig:threshold-incentives}, then formalize the underlying incentives, and distinguish general market incentives from strategic portfolio behavior.

\subsection{Threshold Incentives}

\Cref{fig:threshold-incentives} illustrates the threshold incentive. In the initial equilibrium~$E_0$, the market clears at price~$p_0$ slightly above~$\widebar{p}$. The proxy deduction applies, so renewable generators receive~$p_0 - \delta$ per unit, earning profit equal to area~A.

Now suppose aggregate supply shifts rightward to equilibrium~$E_1$, where the clearing price~$p_1$ falls below~$\widebar{p}$. Although the price is lower, the deduction no longer applies: renewable generators receive the full price~$p_1$ on all output. Their profit in~$E_1$---area~A+B---strictly exceeds their profit in~$E_0$, despite the lower clearing price. As a result, the load-weighted average price paid to renewables may be higher in~$E_1$ than in~$E_0$.

When the clearing price lies slightly above the threshold, even a modest increase in supply can yield a discrete profit gain: the removal of the proxy deduction---which applies to all renewable output---more than compensates for the revenue lost from a lower price. This creates a region in which the marginal value of additional supply is discontinuously large.

\subsection{Formal Characterization}

To formalize this incentive, consider a market with downward-sloping aggregate demand~$D(p)$ and supply from two types of generators: non-emitting generators with total output~$q^r$, and fossil generators whose costs include emission allowance prices.

When~$p_t \geq \widebar{p}$, the proxy deduction applies and non-emitting profit is
\[
\Pi^r_{\text{high}} = (p_t - \delta_t) \, q^r.
\]
When~$p_t < \widebar{p}$, no deduction applies and non-emitting profit is
\[
\Pi^r_{\text{low}} = p_t \, q^r.
\]

The profit function is therefore discontinuous at~$p_t = \widebar{p}$. A price reduction from~$\widebar{p} + \varepsilon$ to~$\widebar{p} - \varepsilon$ changes non-emitting profit by approximately
\[
\Delta \Pi^r \approx (\delta_t - 2\varepsilon) \, q^r,
\]
which is strictly positive for small~$\varepsilon$ whenever~$\delta_t > 0$. The discrete gain from removing the deduction---equal to~$\delta_t \, q^r$---dominates the continuous loss from the price reduction $2\varepsilon \, q^r$. Near the threshold, even small shifts in supply generate discrete profit gains, creating strong incentives to push the equilibrium price below~$\widebar{p}$.

\subsection{General Market Incentives}

The threshold discontinuity creates incentives that may operate even without strategic behavior or market power, through three channels. Overall, we may observe bunching of electricity prices just below the threshold.

First, \emph{supply expansion}: (owners of) non-emitting generators face strengthened incentives to expand output or capacity, shifting the aggregate supply curve rightward. If this moves the equilibrium price below~$\widebar{p}$, all non-emitting output avoids the deduction.

Second, \emph{bid shading}: lowering bids from above to below the threshold $\widebar p$ may be profitable when these bids push the market-clearing price below $\widebar{p}$. This might mean that profit margins of emitting generators are smaller.

Third, \emph{demand-side contraction}: consumers and storage operators facing the threshold discontinuity may have an incentive to curtail demand in hours when the clearing price lies slightly above the threshold. For example, a storage unit that would otherwise charge during such an hour may find it profitable to defer consumption if doing so helps push the market-clearing price below the threshold, thereby lowering the load-weighted average price it faces. This short-run demand response channel reinforces the supply-side incentives discussed above.

Strategic considerations arise particularly for firms with mixed renewable and fossil portfolios, which can internalize the tradeoff between lower fossil margins and avoided deductions on renewable output. Consider a firm producing~$R$ units with non-emitting generators and $F$ units with emitting. Suppose the market clears at~$p^\ast > \widebar{p}$ absent any strategic adjustment. If the firm can push the clearing price below~$\widebar{p}$, renewable revenue rises from~$(p^\ast - \delta) R$ to~$\widebar{p} \, R$.

Let~$\Delta p = p^\ast - \widebar{p} > 0$. The gain from avoiding the deduction is~$(\delta - \Delta p) \, R$. However, lowering the price also reduces inframarginal revenue on the firm's fossil output. The total profit change is approximately
\[
\Delta \Pi_i \approx
\underbrace{(\delta - \Delta p) R}_{\text{net renewable gain}}
\;-\;
\underbrace{\Delta p \cdot F}_{\text{lost fossil revenue}}.
\]
Decreasing the price $p^*$ below the threshold $\widebar p$ is profitable only if~$\Delta \Pi_i > 0$, that is, when
\begin{align}\label{equ:incentive-to-manipulate}
    \Delta p < \delta \cdot \frac{R}{R + F}.
\end{align}
The incentive is increasing in the renewable portfolio share~$R / (R + F)$: firms with large renewable and small fossil positions gain most from crossing the threshold. Note that the marginal cost of fossil plants does not enter \cref{equ:incentive-to-manipulate}: it may even be attractive to bid below marginal cost on emitting units to gain on non-emitting output.

When no single firm has sufficient price influence, threshold manipulation is unlikely. When residual demand is steep and firms hold large mixed portfolios, bunching of prices just below~$\widebar{p}$ may occur. As we show in \cref{sec:incentives-ramp}, this bunching incentive can be eliminated by phasing in the deduction linearly over an interval~$[\underline{p}, \widebar{p}]$ as described in \cref{sec:parameter-design}. Alternatively, the regulator could randomize the threshold, reducing the predictability of the cutoff and thereby the attractiveness of bunching strategies. Note that bunching, even if it distorts the price distribution around~$\widebar{p}$, still lowers the clearing price relative to the status quo. However, it may increase electricity expenditures relative to the counterfactual without strategic responses, since prices cluster just below rather than above the threshold where the deduction would have applied.

\subsection{Dispatch Efficiency within Regimes}

Away from the threshold, the mechanism preserves dispatch incentives. The deduction is a post-clearing settlement adjustment that does not enter the auction algorithm (apart from the pricing stage). When $p_t\ll\widebar{p}$, the policy is ineffective and the dispatch incentives are therefore as with $\delta_t=0$. When~$p_t \gg \widebar{p}$, each additional unit of renewable output earns~$p_t - \delta_t$, still well above near-zero marginal cost; fossil generators continue to earn~$p_t - c_i$. Neither the cost ranking nor the marginal return to output changes, so dispatch order is preserved within each regime.

Near the threshold, this no longer holds. As shown above, firms may find it profitable to bid emitting units below marginal cost to push the clearing price below~$\widebar{p}$ and avoid the deduction on non-emitting output. The threshold thus introduces bid distortions in a neighborhood of~$\widebar{p}$. While distortions may be profitable for a portfolio bidder, they still decrease the clearing price.  Around the threshold price, extreme bunching incentives can be smoothed out as we discuss in the next section.

\subsection{Incentives under the Linear Ramp}
\label{sec:incentives-ramp}

The bunching incentives analyzed above arise from the payoff discontinuity at a single threshold~$\widebar{p}$. Under the linear phase-in schedule $d(p_t)$ defined in \cref{sec:parameter-design}, the deduction increases continuously over the interval~$[\underline{p}, \widebar{p}]$, and the discontinuity is eliminated.

Within the ramp, the remuneration of non-emitting generators is
\[
\widetilde{p}_t = p_t - \delta \frac{p_t - \underline{p}}{\widebar{p} - \underline{p}},
\]
so the marginal remuneration with respect to the clearing price is
\[
\frac{\partial \widetilde{p}_t}{\partial p_t} = 1 - \frac{\delta}{\widebar{p} - \underline{p}}.
\]
This derivative is constant throughout the phase-in interval and non-negative whenever $\delta \le \widebar{p} - \underline{p}$. In particular, the profit function of non-emitting generators is continuous everywhere: there is no price at which a small supply shift produces a discrete profit gain. The strategic manipulation condition~\eqref{equ:incentive-to-manipulate}---which requires a discrete gain $\delta \cdot R$ from crossing a threshold---therefore ceases to apply.

Intuitively, under the ramp a marginal reduction in the clearing price simultaneously lowers the deduction and the price. These two effects offset each other smoothly, so that there is no region in which pushing the price downward is discontinuously profitable. The three channels identified in \cref{sec:incentives}---supply expansion, bid shading, and demand-side contraction---no longer produce bunching because no single price point serves as a discrete trigger.

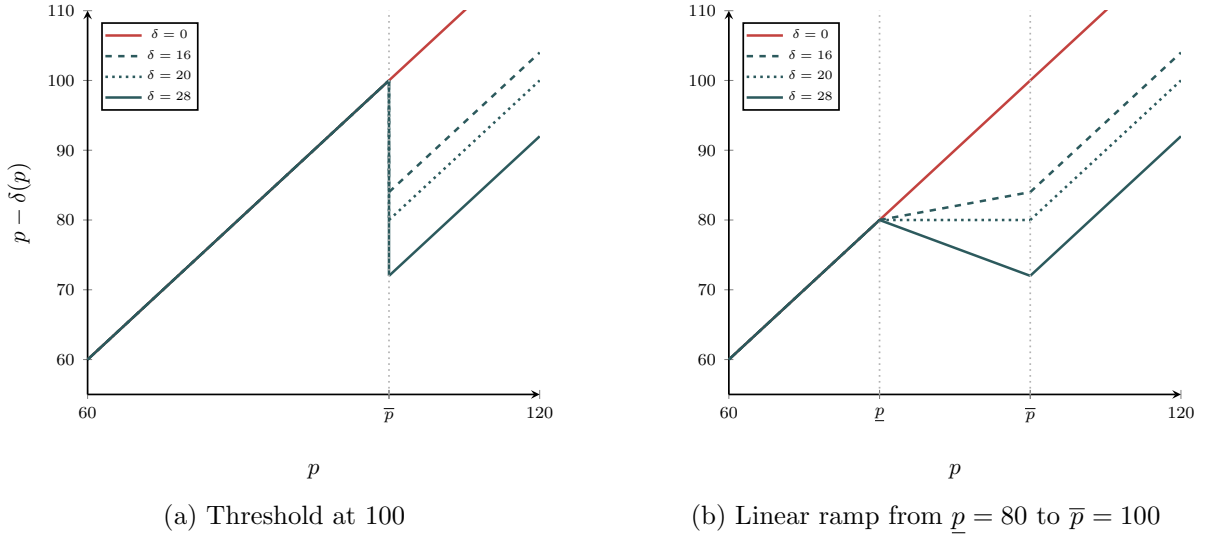
\begin{figure}
    \centering
    \begin{subfigure}[t]{0.47\textwidth}
        \centering
        \resizebox{\linewidth}{!}{%
        \begin{tikzpicture}
        \begin{axis}[
            width=\linewidth,
            height=6.4cm,
            axis lines=left,
            xlabel={$p$},
            ylabel={$p-\delta(p)$},
            xmin=60, xmax=120,
            ymin=55, ymax=110,
            xtick={60,100,120},
            xticklabels={60,$\widebar p$,120},
            tick label style={font=\scriptsize},
            extra x ticks={100},
            extra x tick labels={\phantom{$\widebar p$}},
            extra x tick style={tick label style={yshift=-10pt}},
            domain=60:120,
            samples=200,
            thick,
            scale only axis,
            legend style={font=\tiny, at={(0.03,0.97)}, anchor=north west, inner sep=1pt},
        ]

        \addlegendimage{wifored, very thick}
        \addlegendentry{$\delta=0$}
        \addlegendimage{bluegreen, dashed, very thick}
        \addlegendentry{$\delta=16$}
        \addlegendimage{bluegreen, dotted, very thick}
        \addlegendentry{$\delta=20$}
        \addlegendimage{bluegreen, solid, very thick}
        \addlegendentry{$\delta=28$}

        \addplot[wifored, very thick, forget plot] {x};

        \addplot[bluegreen, dashed, very thick, domain=60:100, forget plot] {x};
        \addplot[bluegreen, dashed, very thick, forget plot] coordinates {(100,100) (100,84)};
        \addplot[bluegreen, dashed, very thick, domain=100:120, forget plot] {x - 16};

        \addplot[bluegreen, dotted, very thick, domain=60:100, forget plot] {x};
        \addplot[bluegreen, dotted, very thick, forget plot] coordinates {(100,100) (100,80)};
        \addplot[bluegreen, dotted, very thick, domain=100:120, forget plot] {x - 20};

        \addplot[bluegreen, solid, very thick, domain=60:100, forget plot] {x};
        \addplot[bluegreen, solid, very thick, forget plot] coordinates {(100,100) (100,72)};
        \addplot[bluegreen, solid, very thick, domain=100:120, forget plot] {x - 28};

        \addplot[gray!60, dotted] coordinates {(100,55) (100,110)};

        \end{axis}
        \end{tikzpicture}
        }
        \caption{Threshold at 100}
        \label{fig:step-remuneration}
    \end{subfigure}
    \hfill
    \begin{subfigure}[t]{0.47\textwidth}
        \centering
        \resizebox{\linewidth}{!}{%
        \begin{tikzpicture}
        \begin{axis}[
            width=\linewidth,
            height=6.4cm,
            axis lines=left,
            xlabel={$p$},
            ylabel={\phantom{$p-\delta(p)$}},
            xmin=60, xmax=120,
            ymin=55, ymax=110,
            xtick={60,80,100,120},
            xticklabels={60,$\underline p$,$\widebar p$,120},
            tick label style={font=\scriptsize},
            extra x ticks={80,100},
            extra x tick labels={\phantom{$\underline p$},\phantom{$\widebar p$}},
            extra x tick style={tick label style={yshift=-10pt}},
            domain=60:120,
            samples=200,
            thick,
            scale only axis,
            legend style={font=\tiny, at={(0.03,0.97)}, anchor=north west, inner sep=1pt},
        ]

        \addlegendimage{wifored, very thick}
        \addlegendentry{$\delta=0$}
        \addlegendimage{bluegreen, dashed, very thick}
        \addlegendentry{$\delta=16$}
        \addlegendimage{bluegreen, dotted, very thick}
        \addlegendentry{$\delta=20$}
        \addlegendimage{bluegreen, solid, very thick}
        \addlegendentry{$\delta=28$}

        \addplot[wifored, very thick, forget plot] {x};

        \addplot[bluegreen, dashed, very thick, domain=60:80, forget plot] {x};
        \addplot[bluegreen, dashed, very thick, domain=80:100, forget plot] {x - 16*(x-80)/(100-80)};
        \addplot[bluegreen, dashed, very thick, domain=100:120, forget plot] {x - 16};

        \addplot[bluegreen, dotted, very thick, domain=60:80, forget plot] {x};
        \addplot[bluegreen, dotted, very thick, domain=80:100, forget plot] {x - 20*(x-80)/(100-80)};
        \addplot[bluegreen, dotted, very thick, domain=100:120, forget plot] {x - 20};

        \addplot[bluegreen, solid, very thick, domain=60:80, forget plot] {x};
        \addplot[bluegreen, solid, very thick, domain=80:100, forget plot] {x - 28*(x-80)/(100-80)};
        \addplot[bluegreen, solid, very thick, domain=100:120, forget plot] {x - 28};

        \addplot[gray!60, dotted] coordinates {(80,55) (80,110)};
        \addplot[gray!60, dotted] coordinates {(100,55) (100,110)};

        \end{axis}
        \end{tikzpicture}
        }
        \caption{Linear ramp from $\underline{p} = 80$ to $\widebar{p}=100$}
        \label{fig:ramp-remuneration}
    \end{subfigure}
    \caption{Remuneration of non-emitting generators under a threshold and a linear ramp deduction}
    \label{fig:linear-ramp-remuneration}
\end{figure}

The phase-in interval should satisfy $\widebar{p} - \underline{p} > \delta$. If $\delta > \widebar{p} - \underline{p}$, the marginal remuneration becomes negative and non-emitting generators would prefer lower clearing prices within the ramp, creating bunching incentives that are similar (but attenuated) compared to the discrete threshold case. At the boundary $\delta = \widebar{p} - \underline{p}$, the remuneration is flat across the phase-in interval: generators are indifferent to price movements within the ramp. \cref{fig:linear-ramp-remuneration} illustrates the muted bunching incentives, for different carbon deductions $\delta$. Note that bunching incentives are most attenuated when $\delta < \widebar{p} - \underline{p}$.

\section{Discussion}\label{sec:discussion}

Several practical objections arise when translating the proposed mechanism from a stylized market model to real-world wholesale electricity markets. We address three issues: institutional implementation (including the identification of generation technologies in order books), the interaction with long-term incentives for renewables, and cross-border price formation in coupled bidding zones.

\subsection{Institutional Implementation}\label{sec:institutional-implementation}

The mechanism admits two equivalent implementations: (i) as an alternative auction design in the day-ahead market, or (ii) a technology-specific levy imposed outside the market with revenue recycled to consumers.

\textit{Auction implementation.} The design of the day-ahead auction can be changed to implement the mechanism. Under the new auction design, the auctioneer determines the market-clearing price $p_t$ as in a standard uniform-price auction. Eligible non-fossil generators receive the carbon cost-adjusted price $\widetilde p_t$, while all other generators receive the market-clearing price $p_t$ in the day-ahead auction. Consumers pay total expenditure as defined in \cref{equ:total-expenditure}. The wedge between $p_t$ and $\widetilde p_t$ is thus embedded directly in wholesale settlement. 

Our binary pricing rule requires identifying which cleared volume comes from eligible non-fossil generation. This is impossible in current European day-ahead auctions: order books contain price--quantity bids, but not unit identity or fuel type at the bidding stage. In addition, many markets---including Austria---use portfolio bidding, so a single bid (function) can aggregate multiple technologies. Cleared bids are therefore not mapped one-to-one to generating units ex ante. This is a technical implementation issue rather than a conceptual problem. To implement the settlement rule in the day-ahead market, bidders must reveal the share of eligible production associated with each price--quantity pair. A practical solution is to adopt unit-based bidding (used, e.g., in Spain and Portugal) instead of the portfolio bidding used in most European countries.

\textit{Implementation as a levy.} Alternatively, the wholesale market is not changed so that all generators receive the uniform market-clearing price $p_t$ in the auction. However, eligible generator $i$ supplying $q_{it}$ in period $t$ pays a levy of $\delta_t q_{it}$ whenever $p_t \ge \widebar p$. The collected revenue is rebated to consumers, for example via network charge reductions or lower electricity taxes. The levy reduces total consumer expenditure by exactly the same amount as the two-price settlement rule, so economic incidence is identical. The difference is purely administrative: the levy leaves wholesale settlement formally unchanged and may be easier to align with existing regulatory and tax infrastructure.

Compared with windfall taxes such as the Austrian Energiekrisenbeitrag--Strom \citep{EKB-S-2022}, the key difference is the treatment of marginal incentives above the trigger price. EKB-S applies a high tax rate $\tau$ to the revenues of electricity generators excluding gas and pump-storage above a statutory benchmark $\hat p$ (e.g., activation around $p_t \ge \hat p$).\footnote{The numbers of $\tau$ and $\hat p$ have changed since 2022. For example, the initial $\tau$ was $0.9$ and later changed to $0.95$. The threshold price $\hat p$ has ranged from an initial 140~\euro/MWh to 90~\euro/MWh.} While the law aggregates over monthly revenue, an hourly stylization leads to a marginal profit of $\hat p + (1-\tau)(p_t-\hat p)$. For sufficiently high~$\tau$, this can weaken renewable output incentives relative to exempt emitting generation. The \cotwo\ proxy deduction instead subtracts a fixed amount in high-price hours: it shifts revenue levels but preserves the full marginal increase in $p_t$, maintaining scarcity-price exposure while still transferring the allowance-cost component embedded in wholesale prices to consumers.

Two further differences are worth emphasizing. First, the proxy deduction is a transparent, rule-based adjustment that can be mechanically indexed to the prevailing allowance price (and activated by an explicit price rule), which makes the redistribution predictable ex ante and avoids ad hoc calibration of ``excess'' revenues. Second, the proposal is designed to reduce electricity expenditure directly: the deducted amount is rebated to electricity consumers (e.g., through lower network charges or electricity taxes), thereby lowering effective prices in the hours in which the mechanism is active. By contrast, windfall taxes typically accrue to the general government budget and need not translate into lower electricity prices unless revenues are explicitly earmarked and recycled back to electricity consumers.

\subsection{Interaction with Long-Term Contracts}

Fundamental incentive effects arise over longer horizons. Inframarginal rents in high-price hours are an important part of the expected return to capital-intensive renewable investments, and the mechanism reduces these rents.

As part of the recent European electricity market reform, two-way Contracts for Difference (CfDs) have been established as the standard support mechanism for new renewable and non-emitting generation. Under the revised Electricity Regulation \citep{EU_Electricity_Market_Design_Reform_PE-1-2024-INIT}, if a member state provides state aid via direct price support for new renewable and nuclear generation, it must use two-way CfDs or schemes that have the same effect. In these contracts a public counterparty and the generator agree on a strike price; when the market price is below the strike price, the public counterparty compensates the generator, and when the market price is above it, the generator pays back the difference. CfDs are intended to provide revenue certainty and reduce investment risk while preserving market responsiveness and avoiding distortions to short-term price signals. Two-way contracts for difference reduce revenue risk and have been increasingly central to European market design reforms \citep{Newbery2016}.

With CfDs in place for new plants, the proposed proxy deduction applies only to existing generators operating purely in the spot market. A new plant under a CfD is exempt from the deduction: its contract settles against the unadjusted uniform clearing price~$p_t$, not against the reduced remuneration price~$\widetilde{p}_t$. Since the proxy deduction is a settlement adjustment that leaves~$p_t$ unchanged (see \cref{sec:market-design-proposal}), the CfD reference price is unaffected and the generator continues to receive the agreed strike price. The two instruments are therefore complementary: CfDs stabilize revenue for investment-grade projects, while the proxy deduction redistributes cyclical inframarginal rents from existing spot-exposed generation during high-price episodes without blunting the carbon price signal in marginal bidding. We propose to apply the proxy deduction to existing spot-exposed, non-emitting generation, and to offer two-way CfD remuneration to new, renewable investments.

\subsection{Cross-Border Price Formation in Coupled Bidding Zones}
European day-ahead markets are coupled across bidding zones through implicit allocation of cross-border transmission capacity. As a result, the marginal generator setting the price in a given zone need not be located within that zone. In particular, a gas-fired peaking plant abroad (but inside the EU ETS agreement) can set the market-clearing price in the domestic zone whenever cross-border transmission capacity is not binding. In such hours, the wholesale price in the domestic zone reflects the foreign plant's marginal cost, including its CO$_2$ component, even though the marginal unit is not subject to domestic settlement rules.

The proposed mechanism should apply to domestic non-emitting generation irrespective of whether the price-setting unit is located domestically or abroad. What matters for the deduction is that the wholesale price exceeds the threshold and thereby embeds an allowance-cost component---not the geographic location of the marginal plant, as long as it is within the ETS zone. The settlement rule is defined over the zonal clearing price and domestic eligible generation; both are observable within the zone regardless of cross-border flows.

A subtler case arises when cross-border transmission capacity is binding and the domestic zone clears at a price above the threshold despite having abundant low-cost renewable generation. If the domestic zone is a net exporter in such an hour, the binding constraint fixes export flows and decouples the domestic price from the importing zone. The price spread between zones constitutes the congestion rent accruing to the TSO. With export volumes fixed at the transmission limit, domestic generators face a locally perfectly inelastic residual demand---domestic load plus the fixed export quantity---and may exercise market power by bidding above marginal cost. In this setting, the proxy deduction serves a dual purpose: it redistributes inframarginal rents to domestic consumers during such episodes, and it dampens the incentive for domestic non-emitting generators to inflate bids against inelastic residual demand. The mechanism therefore remains well-targeted even in the presence of binding interconnector constraints.

\section{Application to Gas Price Shocks}
\label{sec:application-to-gas}

In this section, we discuss how the proposed settlement adjustment can be used as a crisis response to rising gas prices. Under uniform marginal pricing, any increase in the marginal cost of the price-setting technology is transmitted to the market-clearing price and thereby to all inframarginal output. The European gas crisis in 2022 highlighted this mechanism: natural gas prices and \cotwo\ allowance prices were unusually high, which raised wholesale electricity prices and generated large inframarginal rents (``windfall profits'') for non-gas technologies.

This motivates a crisis-oriented variant of the proxy deduction that targets the fossil marginal-cost component embedded in wholesale prices rather than carbon costs alone. Let
\[
  MC_t \;=\; \frac{1}{\eta}\bigl(p_{\text{fuel},t} + e_{\text{fuel}}\, p_{\text{CO}_2,t}\bigr),
\]
denote a proxy for the marginal cost of one MWh of electricity from the reference fossil technology (as in \cref{eq:mc-fossil}), and let
\[
  MC^{\text{ref}} \;=\; \frac{1}{\eta}\bigl(p_{\text{fuel}}^{\text{ref}} + e_{\text{fuel}}\, p_{\text{CO}_2}^{\text{ref}}\bigr)
\]
denote a reference marginal cost level calibrated to pre-shock fuel and allowance prices (e.g., a pre-crisis average or a policy benchmark). We define the excess cost component as the deviation of marginal costs from this reference level:
\[
\delta_t \;=\; \max\{0,\; MC_t - MC^{\text{ref}}\}.
\]
By construction, $\delta_t \ge 0$, so the mechanism never implies a negative deduction when fossil marginal costs fall below the reference level.

To avoid a discontinuous activation at a single cutoff price, we apply the deduction through the linear phase-in $\widetilde p_t = p_t - d(p_t)$ with $d(\cdot)$ as in \cref{eq:linear_ramp}, using $\delta_t$ as the maximum deduction level. Eligible non-emitting generators receive $\widetilde p_t$ while all other generators continue to receive the uniform market price $p_t$.\footnote{The implementing authority must specify the eligible set. In a crisis setting, one natural choice is to apply the deduction to non-gas generation. Whether coal- and lignite-fired units, storage, or subsidized plants should be included depends on relative input costs, the policy objective, and on concerns about incentives and windfall rents.} The lower bound $\underline p$ may be chosen near typical pre-crisis wholesale prices. We let the upper bound vary with the excess-cost component,
\[
\widebar p_t \;=\; \underline p + \varphi\,\delta_t,
\]
where $\varphi>0$ scales the width of the phase-in region. For prices in the ramp region $p_t\in(\underline p,\widebar p_t)$, we have
\[
d'(p_t)=\frac{\delta_t}{\widebar p_t-\underline p}=\frac{1}{\varphi},
\qquad\text{and hence}\qquad
\frac{d\tilde p_t}{dp_t}=1-\frac{1}{\varphi}.
\]
Thus, $\varphi$ directly governs how strongly eligible generators' remuneration responds to changes in the market price within the phase-in interval: larger $\varphi$ implies a wider phase-in region and a slope closer to one, while $\varphi=1$ implies a flat remuneration $\tilde p_t=\underline p$ throughout the interval. We restrict to $\varphi \ge 1$ to ensure $\frac{d\tilde p_t}{dp_t}\in[0,1]$ within the phase-in region.\footnote{The linear phase-in not only mitigates bunching incentives at the activation threshold; it also provides a built-in lower bound for eligible remuneration. Under the ramp schedule with $\varphi\ge 1$, eligible units receive $\widetilde p_t = p_t - d(p_t)$ with $d(p_t)\le \delta_t$, and in particular $\widetilde p_t \ge \underline p$ for all $p_t \ge \underline p$. By contrast, under a discrete rule with a fixed threshold $\widebar p$ and a time-varying deduction $\delta_t$ (e.g., indexed to gas prices), eligible remuneration in activated hours is $\widebar p-\delta_t$. In extreme episodes, $\delta_t$ may become large enough that $\widebar p-\delta_t$ approaches zero or even turns negative. This would create unintended short-run incentives for eligible generators to avoid being settled under the deduction, potentially distorting bidding and participation behavior in high-price hours. One could address this by allowing the activation threshold $\widebar p$ to vary with market conditions (for example, indexing it to a gas-cost benchmark), but such an approach effectively shifts activation to exceptional price spikes and would leave the mechanism largely inactive in moderately elevated-price periods.}

After the deduction, eligible generators receive $p_t-d(p_t)$ per MWh while gas (and other non-eligible) units continue to earn $p_t$ per MWh. The mechanism therefore reallocates part of the inframarginal rent that arises when fossil marginal costs are high, without suppressing scarcity signals or weakening incentives to economize on fossil fuel use. If the deducted revenues are rebated to electricity consumers, the scheme is fiscally neutral and does not require public funds. The incentive effects mirror those discussed in \cref{sec:incentives}.

More broadly, the mechanism implements a rule-based risk-sharing arrangement. Under uniform marginal pricing, fossil-fuel cost shocks are transmitted to electricity consumers through the market-clearing price, even when most generation is non-emitting \citep{EC_Draghi_Report_Competitiveness}. The proxy deduction shifts part of this price risk from consumers to inframarginal producers by reducing eligible generators' revenues in high-price states. Under the linear phase-in, the sensitivity of eligible remuneration to the wholesale price is $\frac{d\tilde p_t}{dp_t}=1-\frac{1}{\varphi}$ within the ramp region, so a fraction $1/\varphi$ of marginal price changes is transferred from eligible producers to consumers. As power systems decarbonize and fossil units set prices only intermittently, such a rule-based mechanism offers an institutional alternative to ad hoc windfall taxes or emergency price caps.

\subsection{Illustrating Counterfactual Savings in 2022}

To illustrate the magnitude of the crisis mechanism, we simulate the expenditure savings that the gas-cost variant of the proxy deduction would have delivered during the European gas crisis of 2022 and compare them to the outcomes of the ``Iberian solution'' in Spain, the only fuel-cost intervention actually implemented during the crisis.

To provide intuition for the cost differences of gas power plants due to higher gas prices, front-month prices on the Title Transfer Facility (TTF), Europe's main gas benchmark, exceeded 300~\euro/MWh$_\text{th}$ at their peak, compared to pre-crisis levels of roughly 20~\euro/MWh$_\text{th}$. Taking a reference gas price of 20~\euro/MWh$_\text{th}$ and an efficiency of $\eta = 55\%$ for a combined-cycle gas turbine, the implied excess electricity cost at peak gas prices was
\[
\delta^{\text{gas}} \approx \frac{1}{0.55} (300 - 20)
\approx 509~\text{\euro}/\text{MWh}
\]
(ignoring the carbon cost differences). Even at a more moderate gas price of 150~\euro/MWh$_\text{th}$, the excess component was $\frac{1}{0.55}(150 - 20) \approx 236$~\euro/MWh. Extreme fuel price shocks thus mechanically generated very large inframarginal rents for non-gas generation whenever gas set the marginal price.

For the counterfactual simulation we calibrate the reference marginal cost to a gas price of 40~\euro/MWh$_\text{th}$ (the cap used in Spain) and the 2021 average EU~ETS carbon price of approximately 53~\euro/t, yielding $MC^{\text{ref}} =$ 92~\euro/MWh. We set $\underline{p}=70$~\euro/MWh to reflect average pre-crisis electricity prices in 2020 and 2021 in Austria and Germany. Moreover, we implement $\varphi=4/3$, so each additional euro in the wholesale price $p_t$ raises eligible generators' remuneration by 25~cents while 75~cents are redistributed to consumers. It follows that $\widebar{p}_t = 70 + \tfrac{4}{3}\,\delta_t$. The maximal proxy deduction is $\delta_t = \max(MC_t - MC^{\text{ref}},0)$.

We apply this mechanism to hourly day-ahead prices in Austria and Germany during July--December 2022, a period in which the Iberian cap was in force. Average net consumer expenditure per MWh falls from 325.5 to 227.3~\euro/MWh in Austria and from 292.5 to 220.8~\euro/MWh in Germany. These are reductions of 30.2\% and 24.5\%, respectively. Eligible generators receive on average 148~\euro/MWh in Austria and 122.4~\euro/MWh in Germany. For comparison, the electricity price in Spain averaged 129.7~\euro/MWh over the same period under the Iberian solution \citep{Fabra-Leblanc-Souza-2025}. The redistribution channel alone therefore delivers sizable consumer relief without distorting marginal costs. Headline price reductions remain smaller than in Spain because the proxy deduction applies only to the eligible (non-gas) portfolio, i.e.,~fossil generators continue to recover their full operating costs, whereas the Iberian mechanism directly subsidized gas-fired generation to suppress the clearing price.

\begin{figure}[htbp]
  \centering
  \includegraphics[width=0.85\textwidth]{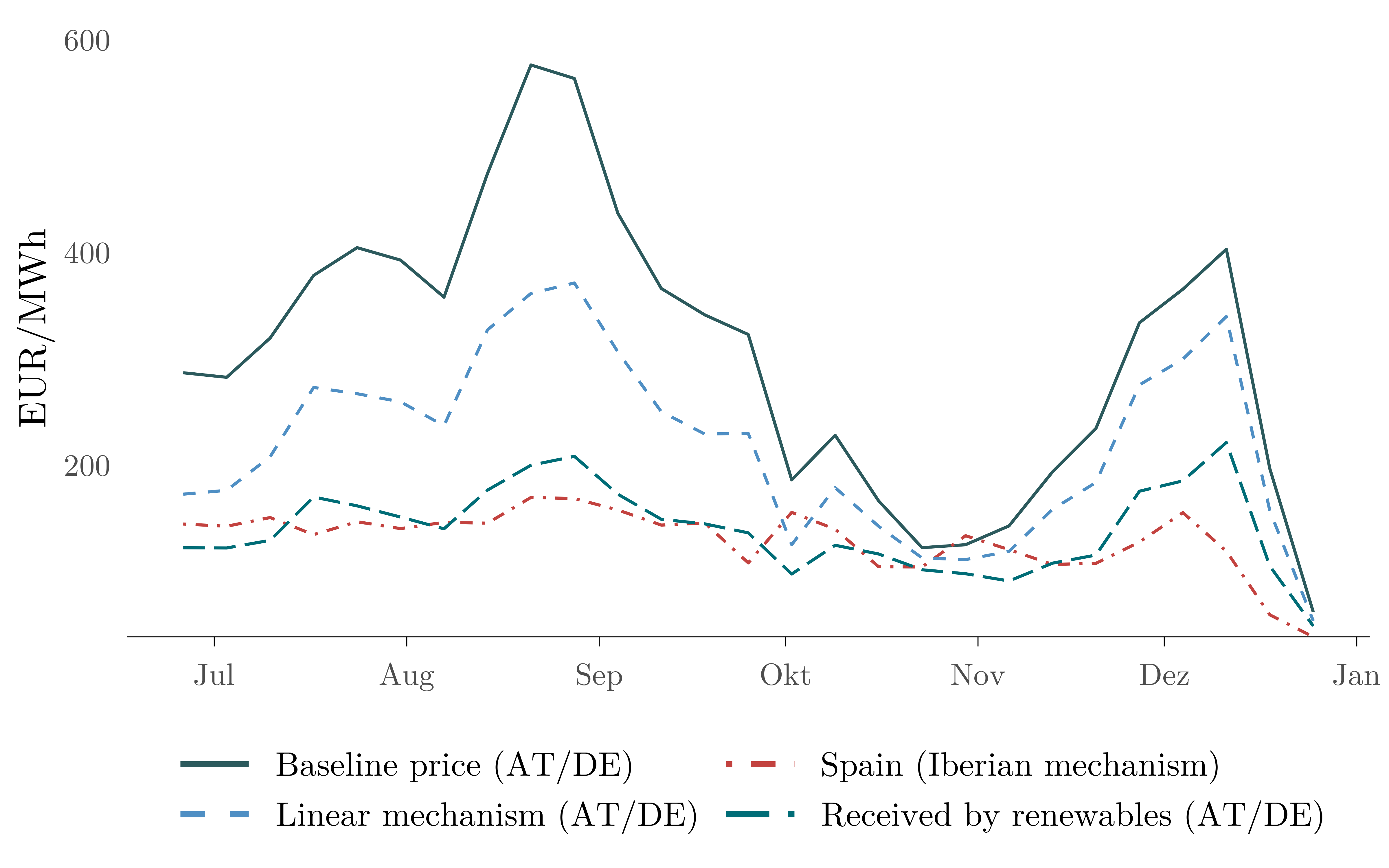}
  \caption{Weekly average wholesale price, eligible-generator remuneration, and net consumer expenditure under crisis mechanisms, July--December 2022}
  \label{fig:p_iberian_comparison_v2}
  \vspace{10pt}
  \begin{minipage}{0.9\textwidth}
    \footnotesize{\emph{Note:} The reference marginal cost is calibrated using a gas price of 40\,\euro/MWh$_\text{th}$ (the Iberian cap) and the 2021 average carbon price. On days where the realized gas price exceeds this reference, the resulting marginal cost differential is subtracted from the day-ahead price for non-gas generators (dot-dashed line). The dashed line shows the resulting average consumer price. Spain's prices reflect actual outcomes under the Iberian mechanism over the same period, following \citet{Fabra-Leblanc-Souza-2025}.}
  \end{minipage}
\end{figure}

\section{Conclusion}\label{sec:conclusion}

European electricity markets face a structural tension. Renewable generation has near-zero marginal costs, yet under uniform marginal pricing, wholesale electricity prices in hours when fossil plants are marginal reflect carbon allowance and fuel costs that non-emitting generators do not incur. The resulting inframarginal rents raise consumer expenditures, weaken electrification incentives, and put pressure on industrial competitiveness.

This paper proposes a settlement rule that addresses this tension directly. A conditional \cotwo\ proxy deduction is applied to the remuneration of non-emitting generators in high-price hours, redistributing part of their inframarginal rents to consumers. The mechanism preserves the uniform clearing price, maintains marginal dispatch and fuel-switching incentives for fossil generators, and leaves the carbon price signal intact. It thereby separates the distributional channel---who captures the rent created by carbon pricing---from the allocative channel that makes carbon pricing effective.

A static accounting exercise using hourly data yields expenditure reductions of about 8.5\% for Austria and 4.7\% for Germany, concentrated in peak-load hours when fossil generation sets the price. A gas-cost variant applied to the second half of 2022 would have reduced average electricity expenditure by about 30\% in Austria and 25\% in Germany, delivering consumer relief comparable to the Iberian mechanism but without subsidizing fossil generation.

The mechanism involves trade-offs. A hard activation threshold creates a payoff discontinuity that may induce price bunching; phasing the deduction in over a price interval via a linear ramp eliminates this discontinuity and preserves full marginal incentives throughout the transition range. Over longer horizons, reduced inframarginal rents lower the expected return on merchant renewable investment. Two-way Contracts for Difference, now the standard EU support mechanism for new non-emitting capacity, provide a natural complement: new plants operating under CfDs settle against the unadjusted clearing price and are therefore exempt from the deduction, while the proxy deduction redistributes cyclical rents from existing spot-exposed generation. The mechanism can be implemented either as a modified auction settlement or as an equivalent technology-specific levy, and applies within coupled European bidding zones regardless of whether the marginal unit is located domestically or abroad.

More broadly, the proposal offers a rule-based institutional alternative to ad hoc windfall taxes and emergency price caps. By reducing effective electricity costs in high-price hours, it may strengthen demand-side electrification incentives, supporting the broader decarbonization objective that carbon pricing is designed to serve.

\bibliography{lit.bib}
\bibliographystyle{aer}

\newpage
\appendix

\section{Additional Tables}

\begin{table}[htbp]
    \centering
    \caption{Revenue loss by technology under hard threshold of 100\,\euro/MWh}
    \label{tab:revenue_loss}
    \small
    \begin{tabular}{lcccccc}
        \toprule
        & \multicolumn{3}{c}{\textbf{Austria (AT)}} & \multicolumn{3}{c}{\textbf{Germany (DE)}} \\
        \cmidrule(lr){2-4} \cmidrule(lr){5-7}
        \textbf{Technology} & \makecell{\textbf{Gen.}\\\textbf{(TWh)}} & \makecell{\textbf{Rev. Loss}\\\textbf{(Mio. \euro)}} & \makecell{\textbf{Loss}\\\textbf{(\%)}} & \makecell{\textbf{Gen.}\\\textbf{(TWh)}} & \makecell{\textbf{Rev. Loss}\\\textbf{(Mio. \euro)}} & \makecell{\textbf{Loss}\\\textbf{(\%)}} \\
        \midrule
        Biomass              &  0.24 &   3.7 & 15.1\% &  35.9 &  431 & 12.9\% \\
        Geothermal           &  ---  &  ---  &  ---   &   0.20 &   2.4 & 13.2\% \\
        Hydro reservoir      &  3.52 &  59.4 & 15.2\% &   1.29 &  21.7 & 14.8\% \\
        Hydro RoR            & 23.9  & 322   & 13.9\% &  12.8 &  147 & 12.7\% \\
        Other renewable      &  ---  &  ---  &  ---   &   0.60 &   7.2 & 13.3\% \\
        Solar                &  5.60 &  31.2 & 10.7\% &  73.9 &  303 &  8.7\% \\
        Wind offshore        &  ---  &  ---  &  ---   &  26.1 &  253 & 11.2\% \\
        Wind onshore         &  8.29 & 112   & 13.8\% & 107   &  872 & 10.5\% \\
        \bottomrule
    \end{tabular}
    \begin{minipage}{\linewidth}
        \vspace{10pt}
        \footnotesize{\emph{Note:} Values represent percentage reduction in revenue relative to the baseline. “---” indicates the technology is absent or negligible ($<0.01$\,TWh) in the respective zone.}
    \end{minipage}
\end{table}

\begin{table}[htbp]
    \centering
    \caption{Sensitivity analysis: threshold and deduction (no pumped storage)}
    \label{tab:sensitivity}
    \small
    \begin{tabular}{cc cc cc cc}
        \toprule
        & & \multicolumn{2}{c}{\textbf{Deduction 23\,\euro/MWh}}
          & \multicolumn{2}{c}{\textbf{28\,\euro/MWh}}
          & \multicolumn{2}{c}{\textbf{33\,\euro/MWh}} \\
        \cmidrule(lr){3-4} \cmidrule(lr){5-6} \cmidrule(lr){7-8}
        \makecell{\textbf{Threshold}\\\textbf{(\euro/MWh)}} & \makecell{\textbf{Base}\\\textbf{Price}} 
          & \makecell{\textbf{New}\\\textbf{Price}} & \makecell{\textbf{Reduction}\\\textbf{(\%)}} 
          & \makecell{\textbf{New}\\\textbf{Price}} & \makecell{\textbf{Red.}\\\textbf{(\%)}} 
          & \makecell{\textbf{New}\\\textbf{Price}} & \makecell{\textbf{Red.}\\\textbf{(\%)}} \\
        \midrule
        \multicolumn{8}{l}{\textbf{Austria (AT)}} \\
        \midrule
         80 & 104.4 & 93.0 & 10.9 & 90.6 & 13.2 & 88.1 & 15.6 \\
         90 & 104.4 & 94.9 &  9.10 & 92.8 & 11.1 & 90.7 & 13.1 \\
        100 & 104.4 & 97.1 &  6.99 & 95.5 &  8.52 & 93.9 & 10.1 \\
        110 & 104.4 & 99.0 &  5.17 & 97.8 &  6.32 & 96.6 &  7.47 \\
        \midrule
        \multicolumn{8}{l}{\textbf{Germany (DE)}} \\
        \midrule
         80 & 92.8 & 86.1 &  7.22 & 84.6 &  8.84 & 83.2 & 10.3 \\
         90 & 92.8 & 87.7 &  5.50 & 86.6 &  6.68 & 85.5 &  7.87 \\
        100 & 92.8 & 89.2 &  3.88 & 88.4 &  4.74 & 87.6 &  5.60 \\
        110 & 92.8 & 90.3 &  2.69 & 89.8 &  3.23 & 89.2 &  3.88 \\
        \bottomrule
    \end{tabular}
    \begin{minipage}{\textwidth}
        \vspace{10pt}
        \footnotesize{\emph{Note:} Prices in \euro/MWh.}
    \end{minipage}
\end{table}

\begin{table}[htbp]
    \centering
    \caption{Policy comparison of linear ramp to hard threshold}
    \label{tab:linear_ramp_results}
    \small
    \begin{filecontents*}{linear_ramp_table.csv}
variant_id,variant_label,zone,policy,total_load_mwh,total_exp_base,total_tax_rev,total_exp_new,avg_price_base,avg_price_new,reduction_abs,reduction_pct
renew_cols,No pumped storage,AT,HT,59.1726098,6178,528,5649,104.4,95.47,528,8.55
renew_cols,No pumped storage,AT,LR,59.1726098,6178,731,5446,104.4,92.04,731,11.84
renew_cols,No pumped storage,DE,HT,465.650611196775,43207,2037,41170,92.79,88.41,2037,4.71
renew_cols,No pumped storage,DE,LR,465.650611196775,43207,3235,39973,92.79,85.84,3235,7.49
\end{filecontents*}
\csvreader[
        tabular=llrrrrrr,
        table head=\toprule \textbf{Zone} & \textbf{Policy} & \makecell{\textbf{Expenditure}\\\textbf{Base}} & \makecell{\textbf{Absolute}\\\textbf{Red.}} & \makecell{\textbf{Expenditure}\\\textbf{New}} & \makecell{\textbf{Reduction}\\\textbf{(\%)}} & \makecell{\textbf{Base}\\\textbf{Price}} & \makecell{\textbf{New}\\\textbf{Price}} \\\midrule,
        table foot=\bottomrule
    ]{linear_ramp_table.csv}{
        zone=\zone,
        policy=\policy,
        total_exp_base=\expbase,
        reduction_abs=\redabs,
        total_exp_new=\expnew,
        reduction_pct=\redpct,
        avg_price_base=\pricebase,
        avg_price_new=\pricenew
    }
    {\zone & \policy & \expbase & \redabs & \expnew & \redpct & \pricebase & \pricenew}
    \begin{minipage}{\textwidth}
      \vspace{10pt}
        \footnotesize{\emph{Note:} HT = Hard threshold 100\,\euro/MWh; LR = Linear ramp $72 \to 100$\,\euro. $\delta = 28$~\euro/MWh. Expenditure figures in millions of euros. Prices in \euro/MWh.}
    \end{minipage}
\end{table}

\end{document}